\documentclass[aps,prb,reprint]{revtex4-1}
\usepackage{amsmath}
\usepackage{ulem}
\usepackage{xcolor}
\usepackage{breakurl}
\usepackage{mciteplus}
\usepackage{graphicx}
\usepackage[colorlinks=true,linkcolor=blue,citecolor=blue,urlcolor=blue]{hyperref}
\usepackage{xr}

\DeclareMathOperator*{\argmin}{arg\,min}

\begin{document}

\author{Miha Gunde}
\affiliation{LAAS-CNRS, Universit{\'e} de Toulouse, CNRS, 7 avenue du Colonel Roche, 31031 Toulouse, France}
\affiliation{CNR-IOM, Democritos National Simulation Center, Istituto Officina dei Materiali, c/o SISSA, via Bonomea 265, IT-34136 Trieste, Italy}
\email{miha.gunde@gmail.com}

\author{Nicolas Salles}
\affiliation{CNR-IOM, Democritos National Simulation Center, Istituto Officina dei Materiali, c/o SISSA, via Bonomea 265, IT-34136 Trieste, Italy}

\author{Anne Hémeryck}
\affiliation{LAAS-CNRS, Universit{\'e} de Toulouse, CNRS, 7 avenue du Colonel Roche, 31031 Toulouse, France}

\author{Layla Martin-Samos}
\affiliation{CNR-IOM, Democritos National Simulation Center, Istituto Officina dei Materiali, c/o SISSA, via Bonomea 265, IT-34136 Trieste, Italy}
\email{marsamos@iom.cnr.it}

\title{IRA: A shape matching approach for recognition and comparison of generic atomic patterns}

\begin{abstract}
We propose a versatile, parameter-less approach for solving the shape matching problem, specifically in the context of atomic structures when atomic assignments are not known a priori. 
The algorithm Iteratively suggests Rotated atom-centered reference frames and Assignments (Iterative Rotations and Assignments, IRA). 
The frame for which a permutationally invariant set-set distance, namely the Hausdorff distance, returns minimal value is chosen as the solution of the matching problem.
IRA is able to find rigid rotations, reflections, translations, and permutations between structures with different numbers of atoms, for any atomic arrangement and pattern, periodic or not. 
When distortions are present between the structures, optimal rotation and translation are found by further applying a standard Singular Value Decomposition-based method. 
To compute the atomic assignments under the one-to-one assignment constraint, we develop our own algorithm, Constrained Shortest Distance Assignments (CShDA).
The overall approach is extensively tested on several structures, including distorted structural fragments. 
Efficiency of the proposed algorithm is shown as a benchmark comparison against two other shape matching algorithms.
We discuss the use of our approach for the identification and comparison of structures and structural fragments through two examples: a replica exchange trajectory of a cyanine molecule, in which we show how our approach could aid the exploration of relevant collective coordinates for clustering the data; and an SiO$_2$ amorphous model, in which we compute distortion scores and compare them with a classical strain-based potential. The source code and benchmark data are available at \url{https://github.com/mammasmias/IterativeRotationsAssignments}.

\end{abstract}

\maketitle

\section*{NOTE:}
This document is the unedited Author’s version of a Submitted Work that was subsequently accepted for publication in Journal of Chemical Information and Modeling, copyright © American Chemical Society after peer review. To access the final edited and published work see \url{https://pubs.acs.org/articlesonrequest/AOR-JCMTX7YW5ZPBSE58C75Q}.

\section{Introduction}
\label{sec:intro}
Shape matching is the ability to find the transformation that best matches a set of points to another set of points. 
In the context of atomic structures, shape matching techniques are exploited in a broad variety of applications, ranging from computer-aided drug discovery\cite{Pharma_Leach2010, Pharma_Giangreco2013, Pharma_Brown2019}, to global structure optimization approaches, such as  genetic-algorithm \cite{GA_Schonborn2009, GA_SIERKA2010, GA_Weal2021} and Basin-hopping Monte-Carlo \cite{HB_Ferr2008, HB_Yang2021}.

Formally, two sets of vector elements are considered congruent or equivalent if they are related by a transformation that preserves distances, i.e. isometric transformation.
Such transformations are rigid translations, rigid rotations, reflections, and permutations of indistinguishable vectors.
The isometric transformation that fulfills the congruence relation between two structures gives a solution to the shape matching problem. This problem can be addressed from different perspectives. In the following, it is stated as an optimization problem. 

If sets $A$ and $B$ represent two atomic structures, e.g. two sets of atomic positions, the congruence relation between them can be written as:
\begin{equation}
    P_B B = {\bf R}A + {\bf t}
    \label{eq:pb3}
\end{equation}
where $P_B$ is a permutation matrix of atomic indices, ${\bf R}$ is a transformation corresponding to either rigid rotation, reflection, or combination of both, and ${\bf t}$ is a translation vector. 

The problem of finding $P_B$, ${\bf R}$, and ${\bf t}$ that best matches one structure to another can be reformulated as an optimization problem:
\begin{equation}
    \argmin_{\bf R,t} \big\{ D({\bf R}A + {\bf t}, B)\big\},
    \label{eq:opt3}
\end{equation}
in which $D$ is a general distance function between two sets, that is i) variant under ${\bf R}$ and ${\bf t}$, ii) invariant under permutation $P_B$, and iii) returns value $0$ when ${\bf R}$ and ${\bf t}$ are such that Eq.~(\ref{eq:pb3}) is satisfied, i.e. when the best match is found. 
It is important to highlight that $D$ does not rely on an internal structural description (encoding), but rather it directly compares the "raw" state of the two structures, since ${\bf R}$ and ${\bf t}$ depend on their relative reference frames.
When distortions and/or deformations are present, the transformation that minimizes Eq.~(\ref{eq:opt3}), does not strictly return a $0$ distance, but some minimum value. 
In that case, the relation between $A$ and $B$ is called a near-congruence, and the isometric transformation ${\bf R}$ and ${\bf t}$ is formally referred to as a near-isometry. 
This minimum distance value provides a measure of the quality of the congruence, i.e. a measure of the similarity between the structures. 
Beyond near-isometry, it is not straightforward to assign a meaning to the distance and transformation that is returned from the optimization of Eq.~(\ref{eq:opt3}). Therefore, a similarity measure obtained from shape matching cannot be thought of only and strictly as a generic similarity metric for arbitrary structures.

\begin{figure}[b!]
    \includegraphics[width=\columnwidth]{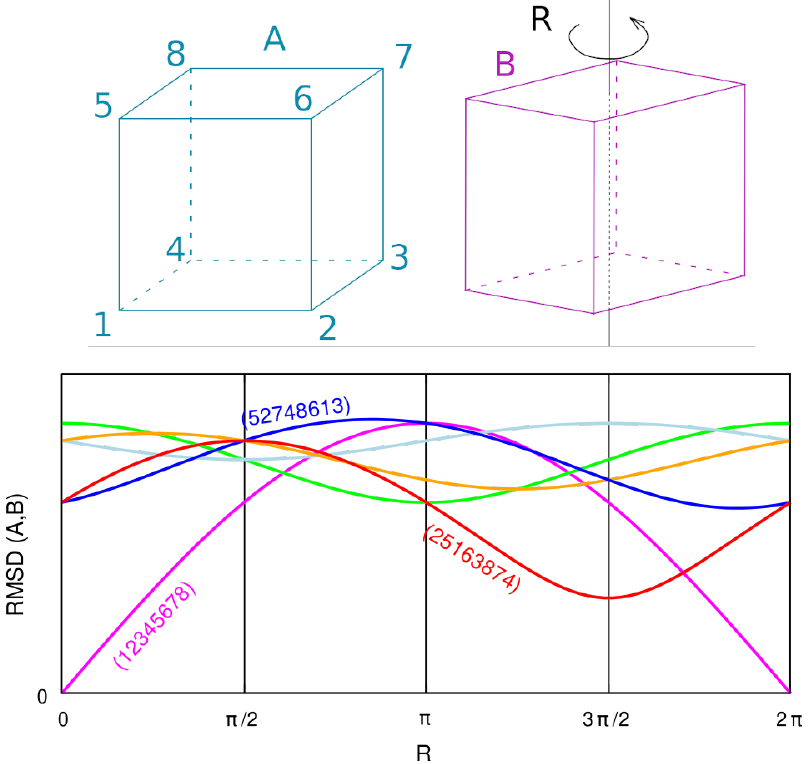}
    \caption{$RMSD$ as a function of rotations ${\bf R}$ and permutations between two identical cubes $A$ and $B$, shown above the plot. Cube $A$ is fixed while $B$ is rotated around the $z$-axis only, for simplicity. Each color in the plot represents a different permutation of the rotated cube, some of them are explicitly labelled. Not all permutations are pictured, as there are in total $N_P= 8! = 40320$ possibilities.}
    \label{fig:cub}
\end{figure}

A widely used set-set distance function, in particular in computational (bio)-chemistry, is Root-Mean-Square-Deviation ($RMSD$), which is usually defined as:
\begin{equation}
    RMSD(A,B) = \sqrt{\frac{1}{N} \sum_i^N{d(a_i,b_i)^2} }
    \label{eq:rmsd}
\end{equation}
where $N$ is the number of points, and $d(a_i,b_i)$ denote an Euclidean distance between points $a_i\in A$ and $b_i\in B$. 
It can immediately be noted that Eq.~(\ref{eq:rmsd}) depends on the ordering of points $i$ in the two sets, its value depends on the permutation $P_B$. 
In other words, $RMSD$ depends on atomic assignments, i.e. which atom from one structure is assigned to which atom from the other structure.
In addition, if we cast the matching problem as finding a global minimum in the phase space of rotations, reflections, and permutations (neglecting for a moment the translations), the definition of $RMSD$ in Eq.~(\ref{eq:rmsd}) does not guarantee the existence of a single connected path from an arbitrary point to the global minimum.
For an example see Fig.~\ref{fig:cub}: a change in the permutation of atoms can lead to a discontinuous jump in $RMSD$ value. 
For this reason $RMSD$ is not directly suitable for a shape matching problem as formulated by Eq.~(\ref{eq:opt3}). In Refs.~\citenum{trott2010} and ~\citenum{griffiths2017}, authors suggested a re-definition of $RMSD$ based on shortest distances, as an attempt to obtain a permutationally invariant quantity. 
Ref.~\citenum{helmich2011} noted that $RMSD$ draws a picture of similarity in an averaging fashion, and proposed an additional criterion for similarity based on the maximal deviation for any atom of $A$ with respect to that atom in $B$. 
Despite the Eq.~(\ref{eq:opt3}) providing stringent criteria for choosing distance functions, in practice there is always some arbitrariness in the choice.

Approaches for finding rotations when the atomic assignments are known and the two structures have the same number of atoms are well established. 
Generally they rely on symmetrization of a special matrix, or minimization of a cost function.
\footnote{ Symmetrization or minimization algorithms for rotations require square matrices. 
As such, the rotations can not be found if the structures have a different number of atoms, without a pre-processing.}
Examples of the two ideas include Lagrange multiplier method \cite{green1952}, matrix symmetrization \cite{strauss1970,fabri2014}, decomposition of a matrix into orthonormal and positive semidefinite matrices \cite{horn1988}, Singular Value Decomposition (SVD) \cite{cliff1966, kabsch1976, arun_svd}, and quaternion eigensystem problem \cite{horn1987, kearsley1989, kneller1991,krasnoshchekov2014} (a review of quaternions can be found in Ref.~\citenum{flower1999}, and more recently in Refs.~\citenum{coutsias2019, hanson2020}).  
Usually the cost function minimized is the $RMSD$ distance.

Finding the assignments between points of two structures is usually called the Linear Assignment Problem (LAP). 
The most widely used general-purpose LAP algorithm is the Hungarian algorithm \cite{kuhn1955, munkres1957}, however others exist, see for example Ref.~\citenum{jonker1987}.
Briefly, it is a mapping from indices of one set to indices of another set, which minimizes a given cost given in the form of a matrix.
When applied to atomic structures, an atom represents an index of a point, and the atomic structure represents a set of points.
Solving this problem might seem simple, but without the knowledge of any intrinsic relation between the atoms,
the complexity increases very quickly as the total number of possible permutations $N_P$ of indistinguishable vectors (atoms) in a structure grows as $N_P  = \prod_{k=1}^m n_k!$, where $m$ is the total number of different atomic types present, and $n_k$ is the number of atoms of atomic type $k$. 

One can also quickly realize that the optimal assignment or mapping of points depends on the relative rotation of the two structures. 
However, algorithms for finding rotations alone are not able to switch permutations by themselves, while algorithms for finding atomic assignments provide the permutation order that minimizes a distance function at fixed rotation, but are not able to suggest rotations that would further minimize it.

To try to overcome such limitations, some strategies have been proposed and are in use in different communities. 
For instance, the algorithm Iterative Closest Point (ICP) \cite{besl} exploits the idea of self-consistent iteration, where each step combines an assignment procedure and consecutive rotation procedure, until a solution is found. 
However, ICP might remain trapped in local minima of the transformation space \cite{pottmann2006}. 
Local minima are a consequence of structural symmetries, see also Fig.~\ref{fig:cub}.
Authors in Ref.~\citenum{blatov2019} suggested an algorithm in which the space of possible rotations and reflections is discretized into a uniform grid of points. 
For each grid-point ${\bf R}$ the optimal atomic assignment $P_B$ is obtained as the optimal assignment of an inter-structure distance matrix with the Hungarian algorithm\cite{kuhn1955}, which is then used to minimize rotations with SVD\cite{kabsch1976}. 
Such strategy is however difficult to optimize, as the number of grid points is not directly related to any property of the system.
A slightly different approach has been proposed in Ref.~\citenum{helmich2011}, with an atomic-centered grid of approximate rotations, in which the farthest atoms from the center are selected as the basis for aligning the reference frames and to find approximate rotations. The atomic assignments are obtained via finding optimal assignment of the inter-structure distance matrix with the Hungarian algorithm.
The authors in Ref.~\citenum{richmond2004} propose an approach for the alignment of molecules based on ideas from image recognition, which relies on filtering methods to obtain atomic assignments. 
Optimal rotations are later resolved by applying an SVD minimization. 
Alternatively, finding a rough equivalent reference frame 
(or Eckart frame\cite{eckart1935,louck1976}) through, for example principal axes of inertia, might also provide a good-enough rotation for identifying reasonable assignments, see for instance Refs.~\citenum{allen2014, wagner2017, temelso2017}.
The principal axes idea is however not suitable for isotropic or compact structures, and crystalline or bulk environments, since the principal axes might be ambiguously defined due to the symmetry of the structures. Moreover, the computation of principal axes of inertia requires the knowledge of associated weights, {\it i.e.} atomic masses.
A successful Monte-Carlo-based decision scheme for finding the global minimum of $RMSD$ \cite{sadeghi2013} has also been reported.

In this work, we present an alternative and versatile, parameter-less approach that solves the general shape matching problem by finding isometries and near-isometries between two (sub-)structures when the assignment is not known a priori, named Iterative Rotations and Assignments (IRA). 
Isometries and near isometries can be found even in the case of structures with different numbers of atoms and belonging to some periodic lattice.
The proposed algorithm iteratively suggests rotated atom-centered reference frames of one structure, to find an approximate rotation in which the matching to the other structure is best. 
This best match provides the one-to-one atomic assignment, thus the permutation $P_B$. When structural distortions are present between the structures, the optimal rotation ${\bf R}$, is later found via SVD\cite{arun_svd}. 
To avoid the ambiguity in the mitigation of improper rotations in SVD and to enable the matching of mirror structures, reflection symmetries are taken into account by also proposing a reflected configuration at each step of the iteration.    
To assess the matching, our approach exploits a truly permutationally invariant set-set distance function, namely the Hausdorff distance\cite{eiter1997}. 
This distance measure is often exploited in the computer vision community, where the shape matching problem is referred to as {\it point set registry}. 
In our implementation, the Hausdorff distance is evaluated after imposing the one-to-one atomic assignment.

We first test the reliability of our proposed matching approach (Sec.~\ref{sec:tests_cdb}), by applying random rigid transformations and permutations to a range of structures, and then applying the shape matching algorithm to re-find them.
Later, the performances are compared to two other algorithms, namely ArbAlign\cite{temelso2017} and fastoverlap\cite{griffiths2017}. In all benchmarks, IRA performs significantly better.
To test behavior in near-congruent structures, we apply the algorithm to two short finite-temperature Monte Carlo trajectories (Sec.~\ref{sec:tests_LJ}).
We next apply it to match and analyse the distortion of cyanine molecule fragments (Sec.~\ref{sec:cyanine}) along a replica-exchange molecular dynamics trajectory from Ref.~\citenum{sara_cyanine}.
We also discuss the use of Eq.~(\ref{eq:opt3}) as a definition of a similarity relation to blindly identify, compare and analyze local structures or fragments. Such sub-structures can be connected or not, and the larger structure to be matched might or not include lattice periodicity.

\section{Our Approach}
\begin{figure}
    \includegraphics[width=\linewidth]{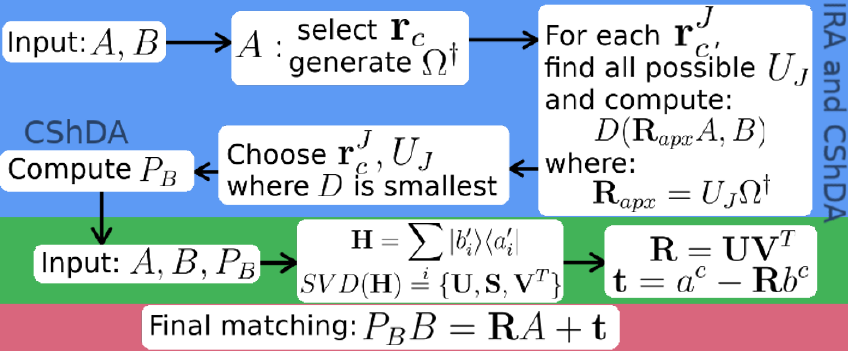}
    \caption{Flowchart of the algorithm. First part of the algorithm colored in blue gives an approximate solution to rotations and translations, and solution to the permutations $P_B$ needed in the second part of the algorithm colored in green, which finds the optimal rotation and translation by utilising the SVD method. Final solution of the matching algorithm is colored in red.}
    \label{fig:flowchart}
\end{figure}

Similarly to other matching techniques briefly summarized in the introduction, we address the general matching problem (Eq.~(\ref{eq:opt3})) in two parts. 
The first part iteratively solves the approximate rotation, which makes it possible to compute the correct atomic assignments. 
The second part uses the atomic assignments to compute the final optimal rotation via standard Singular Value Decomposition (SVD). 
We develop the approach Iterative Rotations and Assignment (IRA, Sec.~\ref{sec:guess}), to obtain the approximate rotation in the first part of our algorithm.
To compute the atomic assignments, we develop our own algorithm: Constrained Shortest Distance Assignment (CShDA, Sec.~\ref{sec:LAP}), that solves the Linear Assignment Problem (LAP) under the one-to-one assignment constraint.
The flowchart representing the full algorithm is shown on Fig.~\ref{fig:flowchart}, where the first part of the algorithm is colored in blue, the second part in green, and the final matching solution is colored in red.

\subsection{Iterative Rotations and Assignment (IRA)} 
\label{sec:guess}
A rigid rotation and translation of a structure by ${\bf R}$ and ${\bf t}$ is equivalent to rotating and translating its reference frame. 
As the distance $D$ in Eq.~(\ref{eq:opt3}) directly compares the "raw" state of the two structures, and ${\bf R}$ and ${\bf t}$ depend on their relative reference frames, the shape matching problem can be addressed as finding a common approximate reference frame between structures $A$ and $B$.
The reference frames that are evaluated in our algorithm are atom centered, with basis vectors chosen as follows.

The atom closest to the geometrical center of $A$ is taken as the central atom and reference frame origin of $A$, i.e. all atoms in $A$ are shifted by the former atomic coordinate vector $-{\bf r}_c$ (in case of periodic structures, periodic boundary conditions are applied). 
Two non-colinear atomic coordinate vectors are subsequently chosen and orthonormalized with the standard Gramm-Schmidt procedure such that ${\bf \hat{e}}_1$ points to an atom. The last reference-frame basis vector is obtained as vector product of the other two, such obtaining a set of three orthonormal basis vectors ${\bf \hat{e}}_1$, ${\bf \hat{e}}_2 $ and ${\bf \hat{e}}_3$. 
The coordinates of $A$ in the new basis can be obtained as: 
\begin{equation}
    A_{ \{ {\bf \hat e} \} }=\Omega^\dagger (A - {\bf r_c}),
\end{equation}
where $\Omega^\dagger$ is the transformation matrix from original reference frame of $A$ to $A_{\{ {\bf \hat{e}} \} }$, formed by the vectors ${\bf \hat{e}}_i$.
To find a similar atom-centered reference frame in $B$, all atoms of the same atomic type as central atom of $A$ are designated candidate central atoms. 
For each candidate central atom $J$, an ensemble of reference frames, and their mirrors are generated by the same procedure as for $A$. 
Namely,
$\{{\bf \hat{e}'}_1$, ${\bf \hat{e}'}_2$, ${\bf \hat{e}'}_3={\bf \hat{e}'}_1 \times {\bf \hat{e}'}_2\}$
and their mirror
$\{{\bf \hat{e}'}_1$, ${\bf \hat{e}'}_2$, ${\bf \hat{e}'}_3={\bf \hat{e}'}_2 \times {\bf \hat{e}'}_1\}$. 
Each candidate central atom $J$ has its atomic vector ${\bf r}_c^J$,
and an ensemble of transformation matrices $U_J$, one for each reference frame guess $\{{\bf \hat{e}}'\}_J$, such that 
\begin{equation}
    B_{\{{\bf \hat e}'\}_J} = U_J^\dagger ( B - {\bf r}_c^J )
\end{equation}
where $U_J^\dagger$ is formed by the vectors ${\bf \hat{e}'}_i$. 

The LAP (Sec.~\ref{sec:LAP}) is solved for all reference frames and central atoms, and 
the combination of reference frame and central atom guess $J$ that return the {\it lowest} set-set distance function $D(A_{\{{\bf \hat e}\}}, B_{\{{\bf\hat e}'\}_J})$, defines permutation $P_B$, the approximate rotation matrix ${\bf R}_{apx}$, and approximate translation vector ${\bf t}_{apx}$:

\begin{eqnarray}
\label{eq:apx}
   {\bf R}_{apx} &=& U_J \Omega^\dagger \\
   {\bf t}_{apx} &=& {\bf r}_c^J - {\bf R}_{apx}{\bf r}_c \notag.
\end{eqnarray}

The distance $D$ is evaluated with the help of our CShDA algorithm, and is equal to the Hausdorff distance, see Sec.~\ref{sec:LAP}, and Sec.~\ref{sec:dh}.

To reduce the number of combinations to be tested in $B$, vectors in $A$ are sorted according to their norm, such that the two atoms taken to generate the basis are as close as possible to the central atom.
The largest norm among these two atomic vectors is taken as a cutoff distance, which is multiplied by a factor (1.2 by default) to account for possible distortions, and taken as maximal-norm threshold for possible basis vectors in $B$. 
The total number of rotations tested $N_R$ thus depends on the compactness (density) and number of nearest neighbors, and goes as $N_R = n_C(n_C-1)$, where $n_C$ are the number of neighbors. 
For a highly compact crystal structure the number of atoms $n_C$ in this sphere can be large (e.g. 15-20), while for molecular structures it is usually much lower (e.g. 5-8). 
The overall order of the procedure is therefore well below $\mathcal{O}(N^3)$, where $N$ is the total number of atoms (see also the Discussion section).
In addition, contrary to the uniform grid proposed in Ref.~\citenum{blatov2019}, our approach does not require blind and massive checks on the number of grid points and their completeness in parsing the rotation space/manifold.

When $A$ and $B$ contain the same number of points/atoms, the search over possible central atoms of $B$ is not required. In that case ${\bf r}_c$ and ${\bf r}_c^J$ is replaced by the coordinates of the geometrical centers of $A$ and $B$, respectively. If any other point that is common to both $A$ and $B$ is known, that particular point can also be used as the center.

If $A$ and (a subset of-)$B$ are exactly congruent, i.e. no atomic deformations, the algorithm would already return the $P_B$, ${\bf R}$, and ${\bf t}$ that exactly minimize Eq.~(\ref{eq:opt3}), as $D({\bf R}_{apx}A+{\bf t}_{apx},B)=0$.

\subsubsection{LAP algorithm: CShDA} \label{sec:LAP}
For the shape matching algorithm presented here, we develop our own atomic assignment algorithm based on shortest distances $d_{ij}$, 
the Constrained Shortest Distance Assignment (CShDA). 
It gives an assignment, or mapping between two atoms $i \rightarrow j$, such that each atom gets a minimum possible cost, under the constraint that each atom can only have one and only one match, so-called one-to-one assignment. 
The idea is that the distances from an atom $a_i\in A$ to all atoms $b\in B$ are used as a cost for computing the assignment of atom $a_i$, such that shortest distances are prioritized for each atom $a_i$ locally.
To showcase, an atom $a_i$ gets assigned an atom $b_j$ with the shortest distance $d(a_i,b_j)$ among all atoms $b$.
However, if during the algorithm an atom $a_i\in A$ is assigned an atom $b_j\in B$ with some distance $d(a_i,b_j)$, and another atom $a_{i'}\in A$ gets assigned the same atom $b_j\in B$ with a distance $d(a_{i'},b_j) < d(a_i,b_j)$, the atom $a_{i'}$ will be prioritized for this $b_{j}$, and the atom $a_{i}$ gets assigned a different atom. 
Symbolically, CShDA iteratively assigns a single atom $a_i \in A$ to a single atom $b_j \in B$ following:
\begin{equation}
   a_i \rightarrow b_j \quad | \quad \min_{b_j \in B} d(a_i,b_j) \quad \forall a_i \in A
    \label{eq:lap_eq}
\end{equation}
with the constraint that $b_j$ has not yet been assigned with a distance lower than $d(a_i, b_j)$, where $d$ is the Euclidean distance between the points.
When applied to a general set of points, this kind of local assignment is sometimes referred to as bottleneck LAP \cite{burkard2009}.

With respect to one of the most widely known general-purpose LAP solvers, the Hungarian algorithm \cite{kuhn1955, munkres1957}, there are two main differences with our proposed CShDA algorithm, explained in the following.

\begin{figure}[htb!]
    \includegraphics[width=\linewidth]{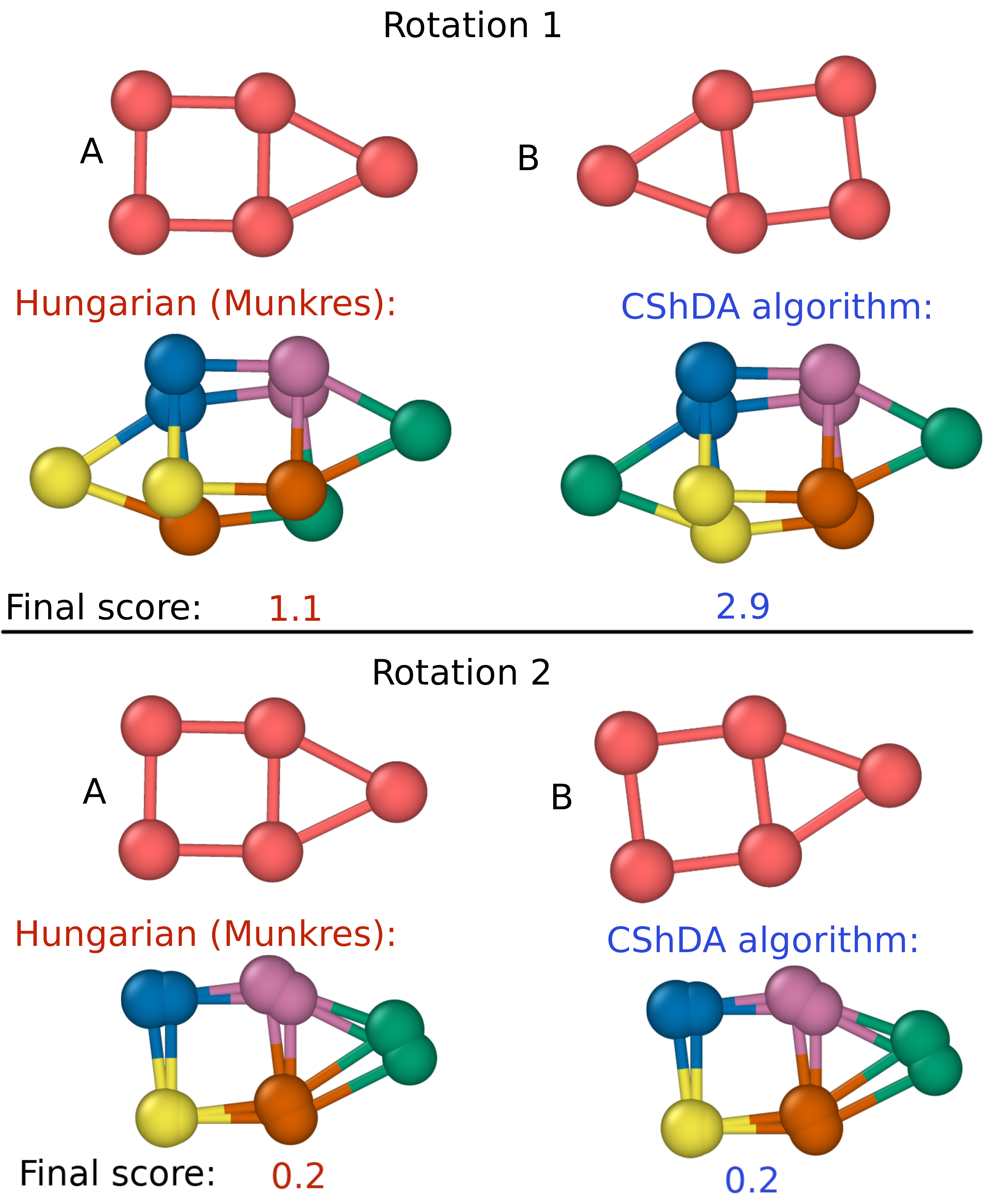}
    \caption{
    A schematic of the assignment problem, solved for structures $A$ and $B$
    in two rotated states.
    On the left the assignment by the Hungarian algorithm following the algorithm proposed by Munkres \cite{munkres1957}, and on the right by our CShDA algorithm. The colors show final assignments of atoms, e.g. a blue atom is assigned to a blue atom, yellow atom to yellow, etc. The final scores are computed as $max( d(a_i,b_i) )$. The first rotated state could represent a particular intermediate step within the iterative rotations procedure (IRA).}
    \label{fig:hungarian_schematic}
\end{figure}

Firstly, the criteria for the assignment of two atoms differ. The Hungarian algorithm assigns indices such that the total sum of the cost is minimized, where the cost of assignment is the distance between two points. In CShDA, each assignment cost is minimized separately, under the one-to-one constraint, where the assignment cost is the distance between points. The CShDA algorithm tends to concentrate the maximum deviations on a small number of atoms, contrary to the Hungarian algorithm that favours smaller deviations, but spread over several atoms. 
Practically, it means that the Hungarian prefers globally "distorted" solutions over rigid single mismatches, see Fig.~\ref{fig:hungarian_schematic}. 

The second difference is that the Hungarian algorithm requires two structures to have equal number of atoms, as the cost of assignment is computed from an all-to-all distance matrix, which needs to be square. While it is true that any square matrix can be made to be non-square by the addition of ghost rows or columns at specific indices, this is not trivial since it is not known a priori which should these indices be. 
Our proposed CShDA algorithm does not have such a constraint. 
The only requirement for CShDA is that the number of atoms $n_A$ in structure $A$ is $n_A \leq n_B$, where $n_B$ is the number of atoms in structure $B$ (this point is also addressed in the Discussion).
In the case when the two sets contain a different number of atoms, there will be some points of $B$ that are left unassigned. 
We enforce that the permutation $P_B$ of set $B$ will in this case be such that the points of $A$ will be assigned to the first $n_A$ points of $P_B B$. 
The unassigned points of $B$ will be permuted to the end of the set. 

\subsubsection{The set-set distance function}\label{sec:dh}
A distance function that fulfils the requirements for solving the shape matching problem as formulated by Eq.~(\ref{eq:opt3}) is the Hausdorff distance function. 

The Hausdorff distance $d_H(A,B)$ between two structures $A$ and $B$ is formally defined as
\begin{equation}
    d_H(A,B) = \max(h(A,B),h(B,A))
    \label{eq:penal_dH}
\end{equation}
where
\begin{equation}
    h(A,B) = \max_{a\in A} \min_{b\in B} d(a,b)
    \label{eq:local_dH}
\end{equation}
where $d(a,b)$ is an Euclidean distance between points $a\in A$ and $b\in B$. 
The value of $h(A,B)$ is the largest value among the smallest distances from points in $A$ to points in $B$.

One can realize that our LAP algorithm corresponds to the $min$ part of the Hausdorff distance in Eq.~(\ref{eq:local_dH}), with the additional constraint of one-to-one assignment. 
The evaluation of Eq.~(\ref{eq:local_dH}) is then the maximal distance $d(a_i,b_i)$ among all points $i$, where the order of atoms $b_i$ follows the assignment provided by the LAP algorithm.

\subsection{Final Optimal Rotation} \label{sec:svd}
In the case in which the two systems are not equivalent, i.e. in the case of near-congruence, after finding the atomic assignments by our IRA algorithm, the optimal rotations are found via an SVD-based algorithm as follows.

Point sets $A$ and $B$ are shifted to their geometrical centers, obtaining $A'=\{a_i' = a_i - a^c\}$ and $B'=\{b_i' = b_i - b^c\}$ where $a^c$ and $b^c$ are the vectors of geometric centers of $A$ and $B$ respectively.
A 3x3 matrix ${\bf H}$ is constructed from $n_A$ points which are common to $A'$ and $B'$ (to enable the decomposition for sets with different number of atoms).
\begin{equation}
    {\bf H} = \sum_i^{n_A} | b_i' \rangle \langle a_i'|,
    \label{eq:hmat1}
\end{equation} 
with $a_i'$ and $b_i'$ the vector points of $A'$ and $B'$, and $|..\rangle \langle.. |$ denoting outer vector product.
The SVD returns three matrices, ${\bf U}$, ${\bf S}$, and ${\bf V}$, such that $SVD({\bf H}) = {\bf USV}^T$, where ${\bf U}$ and ${\bf V}$ are orthonormal matrices corresponding to rotations, and ${\bf S}$ contains the singular values on its diagonal. 
The rotation matrix ${\bf R}$ is then found as 
\begin{equation}
    {\bf R} = {\bf UV}^T,
\end{equation}
and if $det({\bf R})=-1$, then ${\bf R}$ is multiplied by $diag(1,1,-1)$.
The translation vector ${\bf t}$ is found as
\begin{equation}
    {\bf t} = a^c - {\bf R} b^c.
\end{equation}
Rotation ${\bf R}$ and translation ${\bf t}$ found in this way, are such that the $RMSD(A,B)$ is minimized (details on SVD can be found in Ref.~\citenum{arun_svd}).

It is commonly believed that SVD-based algorithms are not particularly suited for matching purposes, due to the ability of SVD to find non-proper rotations \cite{flower1999}, i.e. rotation matrices with negative determinant. Such improper rotations correspond to reflections (sometimes also addressed as pseudorotations \cite{dymarsky2005}), i.e. inversions, or mirroring over some axis, which changes the chirality of a vector set (which is not always desired, e.g.\cite{richmond2004}). 
It has been suggested\cite{arun_svd} to mitigate this issue by multiplying the rotation matrix by $diag(1,1,-1)$, thus forcing a positive determinant. This strategy might however result in a completely wrong rotation, as the matrix ${\bf H}$ depends on the order of points (see Eq.~(\ref{eq:hmat1})).

As our IRA approach (see Sec.~\ref{sec:guess}) suggests permutations corresponding to both rotations and reflections, it is always able to rigorously keep track of what has been suggested, and properly enforce the final rotation matrix to have $det({\bf R})=1$ (corresponding to rotation), or by
multiplying it by $diag(1,1,-1)$ to obtain $det({\bf R})=-1$ (corresponding to reflection). 
Thus consistently providing a correct rotation or reflection matrix.

\section{Results}
\label{sec:tests}

\subsection{Exact congruence and equal number of atoms between sets}
\label{sec:tests_cdb}
\begin{figure}[htb!]
    a)
    \includegraphics[width=0.25\linewidth]{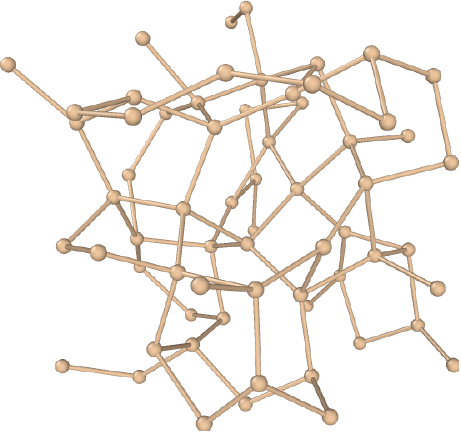}
    b)
    \includegraphics[width=0.25\linewidth]{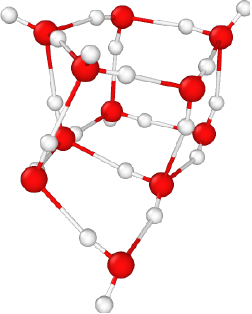}
    c)
    \includegraphics[width=0.25\linewidth]{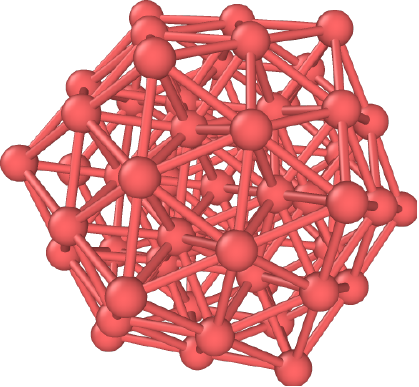}
    \caption{Some sample structures used to test the reliability of the overall algorithm: a) amorphous bulk silicon, b) $n=11$ TIP4P water cluster, and c) $n=52$ Lennard-Jones cluster.}
    \label{fig:ex1}
\end{figure}

The reliability of the algorithm has been first checked by attempting to find the matching between a structure $A$ and a randomized version of that same structure $B$.
The randomized structure $B$ is obtained by randomly permuting, translating by random vector (with norm in the interval (0,10]), rotating by a random angle (in the interval (0,2$\pi$]) along a random rotation axis, and randomly mirroring the structure $A$.
The structures $A$ used for this test are from the Cambridge Cluster Database \cite{cdb}, more specifically we have used the TIP4P water clusters with $n=2$ to $n=21$ molecules of water in the cluster, and the Lennard-Jones (LJ) clusters of sizes $n=3$ to $n=150$ and from $n=310$ to $n=1000$ atoms, from the same database \cite{cdb}. 
We have also used an amorphous Si structure with $n=64$ atoms. Some sample structures are shown in Fig.~\ref{fig:ex1}.
The test is done 10000 times for each of the water cluster structures, 100 times for each LJ structure, and 10000 times for Si structure.
The final matching is evaluated by computing distances $h(A,B)$ and $RMSD(A,B)$ after the matching, they have in all cases both been below the floating point precision value ({\it i.e.,} zero). 
Which implies that with our approach, the correct transformation has always been found without mistake.
The TIP4P test has also been performed by authors in Ref.~\citenum{helmich2011}. 
Their algorithm has failed for $n=10$ once, for $n=11$ once, and for $n=13$ once. 
The same authors reported testing on five amino acids with the same procedure, however the structures of the amino acids claimed to be included in Supporting Information of Ref.~\citenum{helmich2011} have not been found. 

To benchmark IRA with respect to other shape matching approaches, we have performed the same kind of tests against ArbAlign \cite{temelso2017} and fastoverlap \cite{griffiths2017} algorithms. The testing procedure is identical to the previous paragraph, but done on the following datasets. 
From Ref.~\citenum{temelso2017}: datasets of Ne clusters with $n=\{10,50,100,150,200,300,500,1000\}$ atoms, 
water clusters with $n=2$ to $n=21$ and $n=\{25, 40, 60\}$ water molecules, 
FGG peptides with $n=37$ atoms and 4 atomic types, 
and S1-MA-W1 hydrates with $n=17$ atoms and 5 atomic types. 
From Ref.~\citenum{al_dataset}: Al clusters with $n=63$ to $n=160$ in steps of 1, and $n=160$ to $n=310$ in steps of 5 or 10.
From Ref.~\citenum{GaN_dataset}: GaN clusters with $n=12$ to $n=96$ in steps of 2 or 4. 
From Ref.~\citenum{au26_dataset}: Au26 clusters with $n=26$ atoms and a varying number of atoms of a different type. 
From Ref.~\citenum{cdb}: Lennard-Jones clusters with $n=5$ to $n=150$ and $n=310$ to $n=520$.
Each structure from each dataset is tried with 50 random initial transformations, and the final matching is marked as failure if the final distance $RMSD(A,B)$ is greater than threshold 0.001.
The results of this test are reported in Table~\ref{tab:result1}, containing the information on the total number of failures for each dataset. The values of final $RMSD$ distances, for each dataset where failures have occurred, are given in the Supporting Information, in Figs.~\ref{fig:pl_Ne}-\ref{fig:pl_LJ}.

The algorithm ArbAlign\cite{temelso2017} relies on principal axes of inertia as initial guess for rotations, uses the Hungarian\cite{kuhn1955} algorithm for the LAP, and minimizes rotations with SVD \cite{kabsch1976}. It considers 48 pre-defined symmetry operations applied in the reference frame of the principal axes. The algorithm fastoverlap\cite{griffiths2017} is based on kernel correlation. It uses Fourier transform to find maximum correlation between density representations of two structures.

\begin{table}[h!]
    \centering
    \begin{tabular}{l|c||c|c|c}
    Dataset &  $N_s$  & ArbAlign \cite{temelso2017}  &  fastoverlap \cite{griffiths2017}  &    IRA  \\ \hline
    Al~\cite{al_dataset} & 93      & 0/0      & 613/34   &     0/0 \\
    Au26~\cite{au26_dataset} & 6 & 186/4 & *0/0 & 0/0 \\
    FGG~\cite{temelso2017} & 15 & 0/0 & *0/0 & 0/0 \\
    GaN~\cite{GaN_dataset} & 31 & 50/1 & *294/14 & 0/0 \\
    LJ~\cite{cdb} & 357 & 45/1 & 1177/113 & 0/0 \\
    Neon~\cite{temelso2017} & 16     & 100/2    &   82/8   &   0/0 \\
    S1MAW1~\cite{temelso2017} & 20 & 0/0 & *0/0 & 0/0 \\
    water~\cite{temelso2017} & 70 & 0/0 & *217/11 & 0/0
    \end{tabular}
    \caption{Results of the efficiency test of the three algorithms. Each dataset is referred to by its name, $N_s$ is the number of different structures in each dataset. Each structure from each dataset was tested with 50 random initial transformations. The tabulated values are in the form $m/n$, where $m$ is the total number of failures, and $n$ is the number of structures in which the failures occur. Values marked with *: the structures in this dataset include several atomic types, which fastoverlap cannot distinguish. }
    \label{tab:result1}
\end{table}

From the results of our benchmark test in Table \ref{tab:result1}, we can conclude the following. The algorithm ArbAlign \cite{temelso2017} has problems to find the correct rigid transformation in structures where the principal axes of inertia are ambiguous, as anticipated in our introduction. This is very clear from the Au26 dataset from Ref.~\citenum{au26_dataset}, which includes cylindrical shape structures, where only the principal axis along the cylinder is well defined. We note that since each structure was tried 50 times, the result of 186 failures in 4 structures (see Table \ref{tab:result1}) indicates that on average, there were 46 failed attempts out of 50 trials per structure. 

On the other hand, the algorithm fastoverlap \cite{griffiths2017} shows a higher overall rate of mismatches, but with broadly dispersed failures. Interestingly, the final values of distance from fastoverlap show clustering around several distinct values for each structure 
(see Figs.~\ref{fig:pl_Ne}-\ref{fig:pl_LJ} in the Supporting Information), which might be the signature of trapping on some local minima.

Our proposed IRA algorithm shows a success rate of 100\% across all of the structures tested. 
We can say with high confidence that it is fully reliable at finding any rigid transformation between two congruent structures.

\subsection{Near congruence and equal number of atoms between sets}
\label{sec:tests_LJ}

To test the performance under conditions of near congruence, i.e. the structures present some deformations - we perform a short NVT-ensemble Monte Carlo (MC) simulation for a LJ-20 cluster from the Cambridge Database \cite{cdb} at two different temperatures. 
The specific temperatures used are $T=0.02$ and $T=0.3$ in the reduced units. These two values have been chosen as corresponding to "low" and "a bit higher", and are only used to induce some atomic vibration.

\begin{figure}[htb!]
    \includegraphics[width=0.99\linewidth]{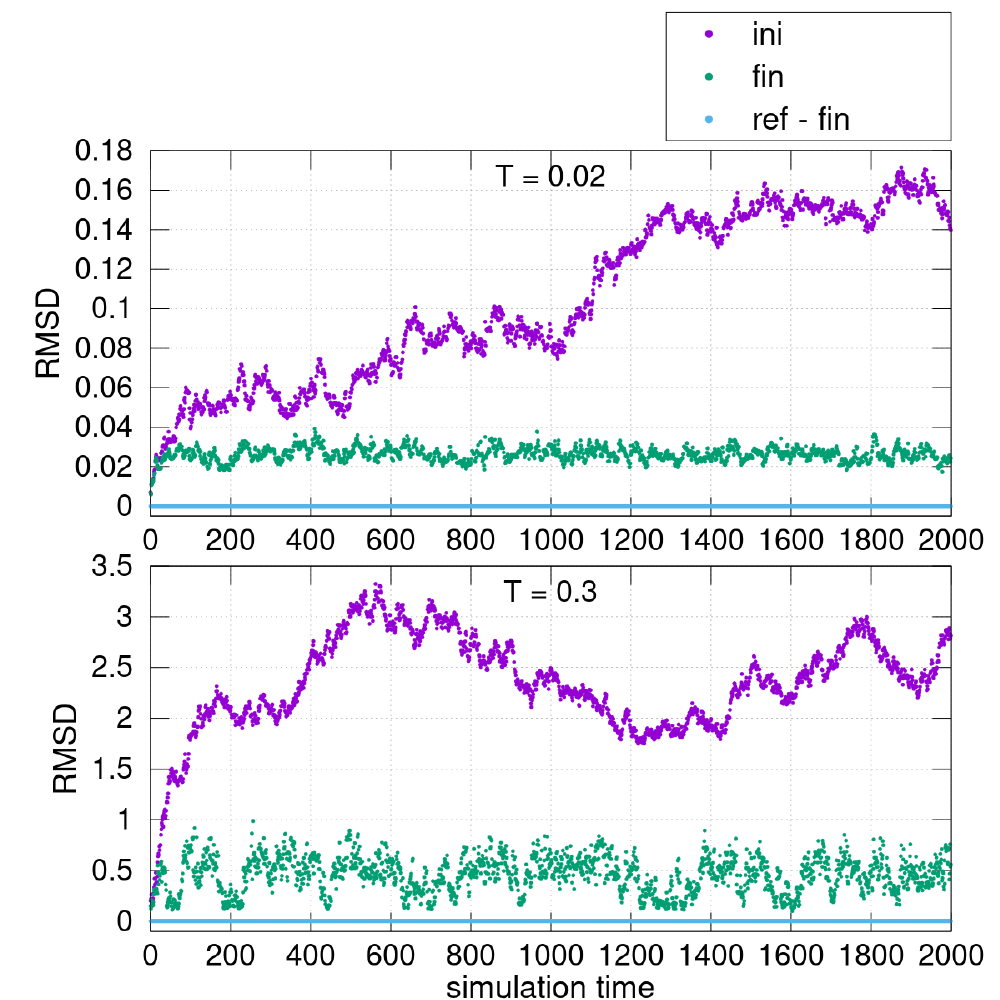}
    \caption{Plot of $RMSD_{ini}$, $RMSD_{fin}$, and the difference $RMSD_{ref}-RMSD_{fin}$
    for temperatures (top) $T=0.02$, and (bottom) $T=0.3$. }
    \label{fig:mclj20}
\end{figure}

We take the equilibrium configuration of the cluster as reference structure $A$. 
At each step of the MC simulation, the current structure is taken as $B$, and the distance $RMSD_{ini}=RMSD(A,B)$ is calculated. 
During the MC, the structure undergoes some distortion, translation, and rotation, but not permutation of atoms. 
We can readily apply the SVD method to obtain rotation that minimizes $RMSD(A,B)$ at current step, store this $RMSD$ value as $RMSD_{ref}$. 
Then apply random rotation, reflection, translation, and permutation to structure $B$, and run our shape matching algorithm on it, to obtain $B'$ aligned to $A$, and calculate distance $RMSD_{fin}=RMSD(A,B')$. The distance $RMSD_{fin}$ should be equal to $RMSD_{ref}$ if our algorithm has successfully found the right transformation. 
The results are shown on Fig.~\ref{fig:mclj20}. The difference $RMSD_{ref} - RMSD_{fin}$ on every step is on the order of floating point precision error ({\it i.e.,} zero), confirming the ability of the presented approach to find the correct matching transformation efficiently. 

The non-zero value of $RMSD_{fin}$, provides with a measure of the congruence between the structures.

\subsection{Near congruence and different number of atoms between the sets}
\label{sec:cyanine}
\begin{figure*}[htb!]
    \centering
    \includegraphics[width=0.7\linewidth]{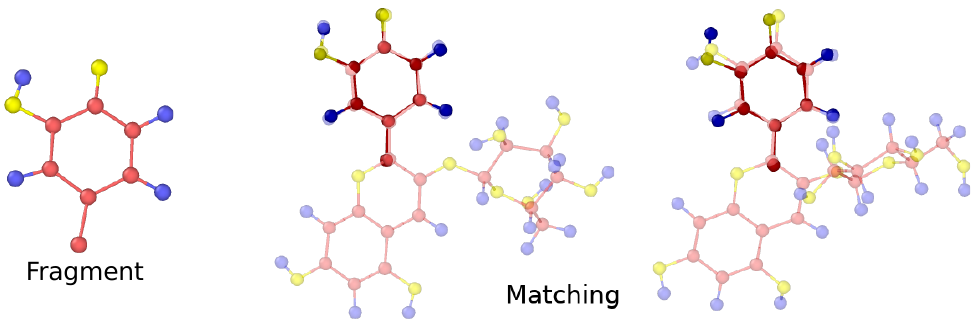}
    \caption{The fragment to be matched, and two instances of the final matching of the molecule, the atoms of the fragment are shown with a darker shade for better distinction. Red, blue and yellow atoms correspond to Carbon, Hydrogen and Oxygen atoms respectively, the same color code is used in the following.}
    \label{fig:cyanine_fragment_match}
\end{figure*}
\begin{figure}[htb]
    \includegraphics[width=\linewidth]{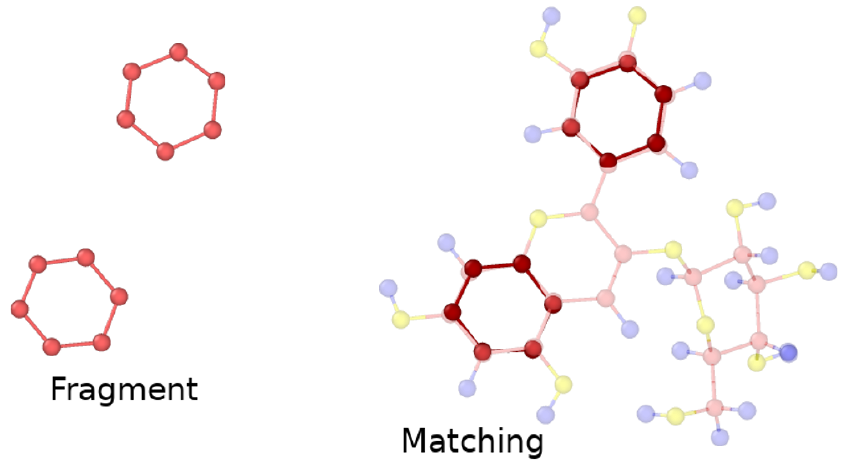}
    \caption{A disconnected fragment, and matching of a molecule.}
    \label{fig:cya_disconnect}
\end{figure}

In order to show the ability and performances of our approach in finding the correct transformation and atomic assignment that best matches the structural fragments to a larger structure, 
we use a trajectory of replica-exchange molecular dynamics simulation of the cynanine molecule (data provided by authors of Ref.~\citenum{sara_cyanine}).
We select two kinds of fragments, a connected one shown in  Fig.~\ref{fig:cyanine_fragment_match}, and a non-connected one shown in Fig.~\ref{fig:cya_disconnect}. 

During the trajectory, the atoms move and distort the molecule, but they do not permute. Thanks to this, we can apply a similar test for reliability as in the previous section.
We choose a fixed reference fragment $A$, and compute the optimal rotation of molecule $B$ using SVD, giving $RMSD_{ref}=RMSD(A,B)$.
Then we randomly rotate, reflect, translate, and permute structure $B$, and run our shape matching algorithm on it, to obtain $B'$ aligned to fragment $A$, and calculate $RMSD_{fin} = RMSD(A,B')$.
The distances $RMSD_{ref}$ and $RMSD_{fin}$ should be equal if the right transformation has successfully been found.
The sum in all $RMSD$ calculations in this case goes up to number $n_A$ of atoms in fragment $A$.

The result when structure $A$ is the connected fragment from Fig.~\ref{fig:cyanine_fragment_match}, is that out of the eighty thousand configurations in the trajectory, there are 313 instances of the difference $RMSD_{ref}-RMSD_{fin}$ being above the floating point precision value. These instances represent structures where the algorithm has mismatched the fragment. 
Some of the reasons for this behaviour are explored in the discussions section (Sec.~\ref{sec:discussion}).
However a deep analysis of the particular instances is beyond the scope of the current paper.

Tracking the number of mismatches when structure $A$ is the non-connected fragment from Fig.~\ref{fig:cya_disconnect} is not straightforward, since the two hexagons do not move rigidly. As a consequence, $RMSD_{ref}$ as defined previously is ambiguous. 

\section{Discussion}
\label{sec:discussion}
\begin{figure}[htb!]
    \includegraphics[width=0.9\linewidth]{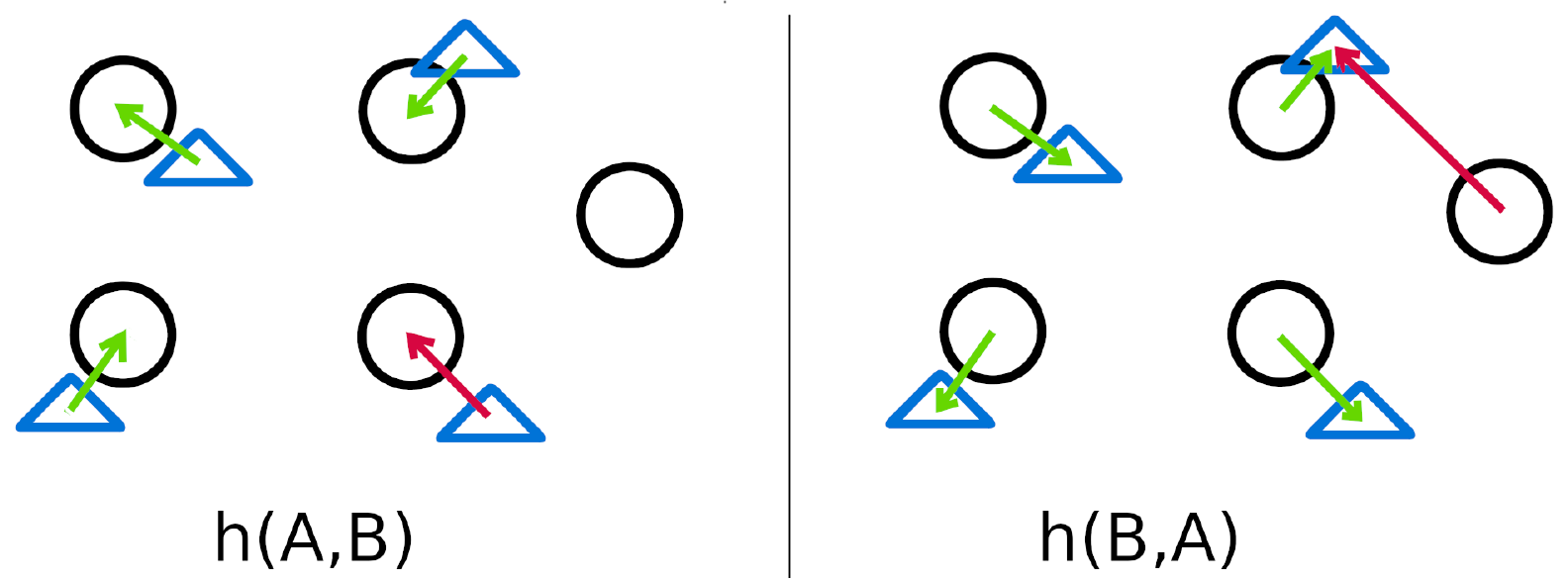}
    \caption{Schematic representation of the difference between $h(A,B)$ on the left, and $h(B,A)$ on the right, when $A$ and $B$ contain different number of points. Set $A$ is represented by triangles, set $B$ by circles. Arrows show the minimum distances between points in green, and the maximum value in red, $h(A,B)$ and $h(B,A)$ respectively.}
    \label{fig:hAB_hBA}
\end{figure}

In the IRA part of the algorithm (Sec.~\ref{sec:guess}), the evaluation of Hausdorff distance $h(A,B)$ is compliant with the one-to-one matching constraint of the CShDA, and strictly corresponds to distance function $D$ in Eq.~(\ref{eq:opt3}).
Due to the relatively low number of atoms in the atomic structure matching, the usage and implementation of the Hausdorff distance needs some attention. 
The expression for $h(A,B)$ in Eq.~(\ref{eq:local_dH}) is only commutative when $A$ and $B$ contain the same number of points, which is the reason the expression for Hausdorff distance is generally written in the form of Eq.~(\ref{eq:penal_dH}), which penalizes the situation where some points are present in one structure but not in the other.
Fig.~\ref{fig:hAB_hBA} schematically shows the shortest distances between points of set $A$(triangles) and points of set $B$(circles) as arrows, where the largest among them is colored in red and represents the value of $h(A,B)$, and $h(B,A)$ respectively. 
As described in Sec.~\ref{sec:LAP}, the assignment of atoms is done under the one-to-one constraint, which poses a problem for the situation of $h(B,A)$ on the right side of Fig.~\ref{fig:hAB_hBA}, where $B$ contains more atoms than $A$, since two atoms of $B$ get assigned to the same atom of $A$.
A mitigation for avoiding this problem is to systematically impose that the number of atoms $n_A \le n_B$, which is the situation of $h(A,B)$ on the left side of Fig.~\ref{fig:hAB_hBA}. 
This imposition also opens up the possibility of matching fragments.
However, the fragment as a whole needs to be a substructure of the larger structure, i.e. our proposed algorithm is not finding the largest common subset of both the structures.

As the value of $h$ only takes the maximal distance in Eq.~(\ref{eq:local_dH}), it only contains information about one specific atom/point.
This particularity can be advantageous in cases of low distortions between the structures, where the value of $h$ is low, meaning that all atoms are within a low-distance $h$ of the reference positions. 
Larger distortions lead to higher $h$ value, which can hide the behavior of any specific atom. 
A high $h$ can be due to single atom distortion, and any information on other atoms is completely obscured. 
This property of Hausdorff distance is often described as high susceptibility to noise.
It opens the possibility of a situation in our algorithm, where a "wrong" assignment gives a transformation $U^{\dagger}_J$ whose distance $D(A_{\{{\bf \hat e}\}}, B_{\{{\bf\hat e}'\}_J})$ is lower than the distance $D$ when the transformation is given by the "correct" assignment, which then leads to a wrong final assignment and transformation.
Replacing the $h$ with a sum of minimal distances, which should capture a more "collective" behaviour of the atoms, has not shown any significant changes in the performances with the highly distorted cyanine molecule tests (Sec.~\ref{sec:cyanine}). 
The mismatches still happen at large set-set distance values. 
The choice of a particular set-set distance function is therefore not crucial, as long as the distance complies with the permutational invariance, and translational and rotational variance, imposed by Eq.~(\ref{eq:opt3}).
The "mismatches" are rather due to attempting to match structures that are far from congruence. 
Which raises the general question for any structure similarity approach, how meaningful can it be to attempt matching such structures, and how could the results be interpreted?
On the specific and known case of the cyanine we were able to assess that there were mismatches, but for huge data sets for which the parsing is generally blind, the meaning of large distances and their interpretation should be of concern.

It is possible to reduce the number of mismatches by assuming some prior knowledge on the system.
The first step of our IRA algorithm (Sec.~\ref{sec:guess}) selects a central atom in structure $A$ by the criteria of closeness to the geometrical center of $A$. 
The second step is to select a basis ${\{{\bf \hat e}\}}$ for a reference frame in $A$, which is based on positions of atoms around the central atom. 
Then the structure $B$ is searched for the equivalent basis ${\{{\bf\hat e}'\}_J}$. 
When large distortions are present in structure $B$, there is no guarantee that the basis found in $B$ is equivalent to the basis found in $A$, or that it even exists. 
If we assume that there still exist local environments in the two structures that are congruent to each other, then the central atom of $A$ could be chosen as the atom for which its local environment is the most similar to any local environment in $B$.
Choosing the central atom in $A$ according to that criterion in the case of cyanine for instance, reduces the number of mismatches by an order of magnitude (313 originally, 30 with this choice).

As already mentioned in Sec.~\ref{sec:guess}, the total number of rotations tested $N_R$ is greatly dependent on the structure. 
In this respect, the Al dataset, along with LJ and Ne datasets from the benchmark test in Sec.~\ref{sec:tests_cdb}, represent worst-case scenarios for IRA as all atoms are of the same atomic type, and the structures are close-packed, which yields the highest number of reference frames to be tested. 
This number is related to the structure surrounding the origin point, as mentioned in Sec.~\ref{sec:guess}. 
For example, in the Al dataset\cite{al_dataset}, the number of rotations tested for each member structure varies on the range [2, 154], without any apparent rule (see also Fig.~\ref{fig:al_nbas} in Supporting Information). 
In that example, there is a single origin point, which is set to the geometrical center of the structures.
A higher number of rotations needs to be tested when the geometrical center coincides with an atomic position. 
In that case, a larger number of atoms is included in the radial cutoff region, which defines the possible reference frames. 
Conversely, when the geometrical center falls in between atoms, the number of atoms in the region is lower, and thus less reference frames have to be tested. 
In the case of matching structures with different number of atoms, the origin point is set by the central atom in structure $A$. 
In that case, each possible central atom of structure $B$ gets tested with a number of rotations that depends on the local environment of that atom.
In any case, the number of rotations tested is not explicitly related to the total number of atoms $N$, but related to the density of atoms in the region around the origin point, and the number of possible origin points. 
When prior knowledge of the origin point in the form of a known central atom is assumed, as discussed in the previous paragraph, the number of rotations tested is given only by the local environment surrounding that specific atom.
The overall performance thus depends on the specific atomic structure, and any prior knowledge influencing the choice of the origin point.

In situations when we know that the two structures being matched are sufficiently similar, the multiplication factor 1.2, used for the cutoff can be reduced, but the value should in any case remain above 1.0.
This effectively reduces the search space of rotations, and the algorithm can be faster as a result. 
When matching structures with different number of atoms, making a computational effort to reduce the number of candidate central atoms, as previously mentioned, can also be very beneficial for the speed of the algorithm, as it reduces the set of possibilities.
In situations where the equality of two structures is being tested with a certain known threshold for equality, heuristic approaches can be used on top of the logic of the IRA and CShDA algorithms, to exit certain loops as soon as certain criteria are met. 
This method has the potential to speed up the algorithm considerably, however at the cost of generality.
Because of the non straight-forward relationship between the speed of IRA algorithm and the atomic structure, a discussion about scalability would hardly be useful. 
As point of reference for the timing, our fortran implementation of IRA as described in this work, running on a single core of a standard laptop: matching the LJ $n=100$ cluster\cite{cdb} with a randomized version of itself takes about 0.02 seconds with 40 rotations tested, and 0.15 seconds for the LJ $n=400$ cluster with 12 rotations tested. 
However these numbers cannot be generalized at all.

Similarly, when matching structures with different number of atoms, the best-case and worst-case scenarios in terms of overall speed of execution, would be the following. 
Best-case would be matching a fragment of a low-density structure, to a slightly larger structure with a small number of possible central atoms, meaning the central atom of $A$ has an atomic type that is not very present in structure $B$ (as is the case for example for some organic compounds).
The worst-case scenario would be matching a fragment of a high-density structure, to a much larger structure with many possible central atoms (as for example in close-packed bulk structures).

Once the transformation that best matches one structure to the other is found, the corresponding set-set distance value becomes a similarity measure or a distortion score: 
a similarity measure that is not an arbitrary choice, but that arises from a minimization. As our approach is also able to match fragments (connected or not), including a lattice periodicity, it can provide with a similarity measure for any part of any structure. 

\begin{figure*}[htb!]
    \centering
    \includegraphics[width=0.8\linewidth]{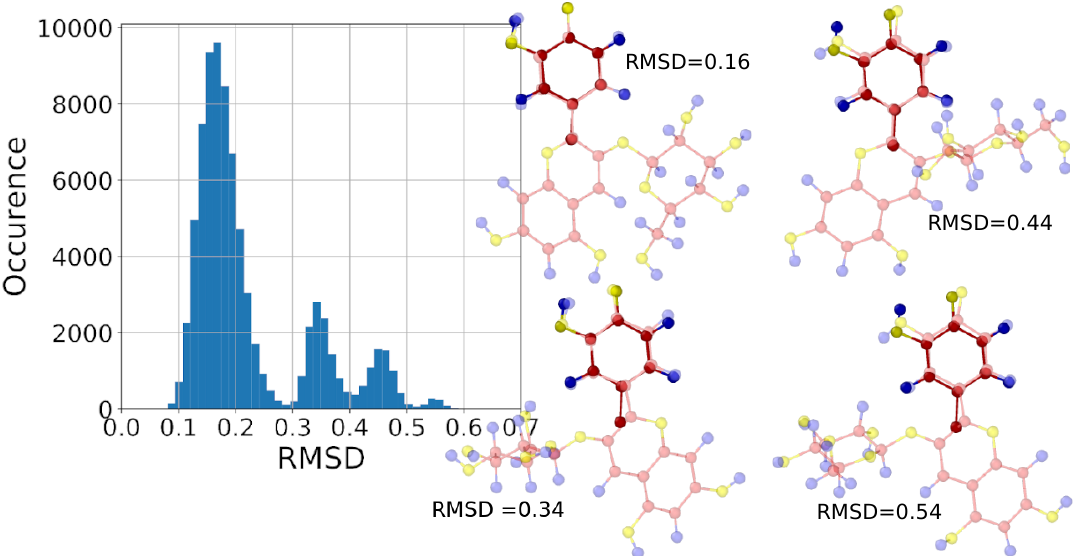}
    \caption{Histogram of $RMSD$ values of the final matching for 80 thousand trajectory steps. We clearly see four peaks, representing four clusters of structures in the MD trajectory, the typical member structure corresponding to each peak is shown. The viewing angle is such that the reference fragment, shown in darker colors, is kept fixed on all images.}
    \label{fig:cya_histo1}
\end{figure*}

\begin{figure}[htb!]
    \includegraphics[width=0.999\linewidth]{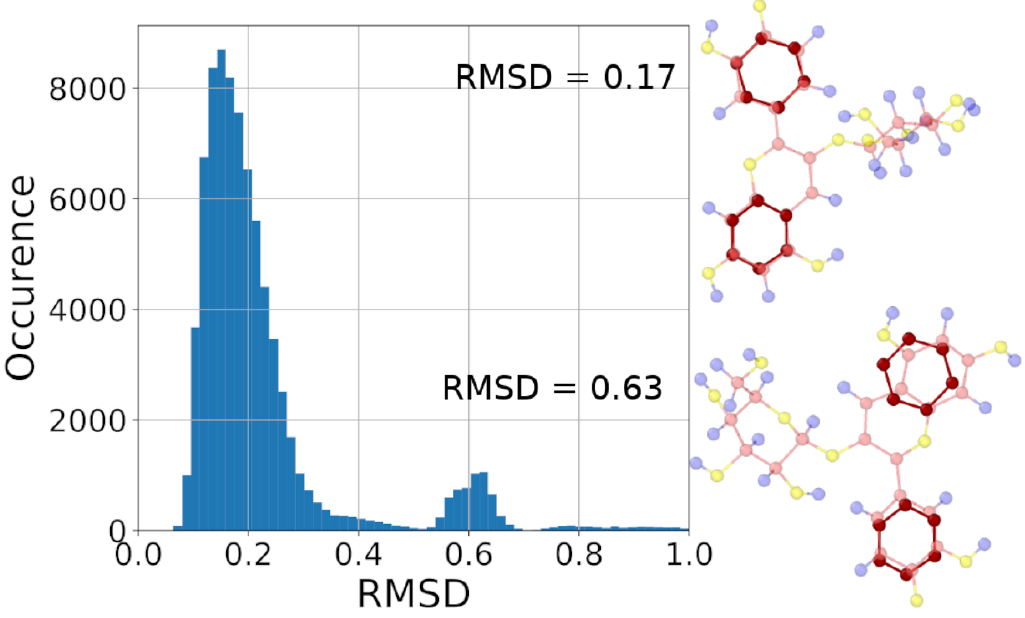}
    \caption{Histogram of $RMSD$ values of the final matching of a disconnected fragment. Two peaks can be identified, corresponding to the grouping of the structures into two clusters. A representative structure from each cluster is shown.}
    \label{fig:cya_histo2}
\end{figure}

Exploited in (semi)-blind fragment exploration, our approach could aid in revealing the most important collective coordinates, which ultimately cluster the data set along the relevant collective axes.
For example, Fig.~\ref{fig:cya_histo1} and Fig.~\ref{fig:cya_histo2} show two sample histograms of $RMSD$ for the final matching of the eighty thousand trajectory steps of the cyanine example (see Sec.~\ref{sec:cyanine}) with respect to two sample reference fragments. The cases of mismatching are excluded from these plots. In Fig.~\ref{fig:cya_histo1}, four 
peaks can be identified, representing the grouping of structures in the MD trajectory into four clusters.
From the representative fragments belonging to each cluster, we can notice that there is a H-atom (blue) that rotates around an O-atom (yellow), and that the rest of the molecule that is attached through the bottom C-atom (red) of the fragment is roughly oriented in two main directions.
Indeed the original paper with the cyanine molecule \cite{sara_cyanine} reports the dihedral angle going through the bottom C-atom as one of the relevant axes which clusters the whole data set into two main groups.

\begin{figure*}[htb!]
    \centering
    a)
    \includegraphics[width=0.4\textwidth]{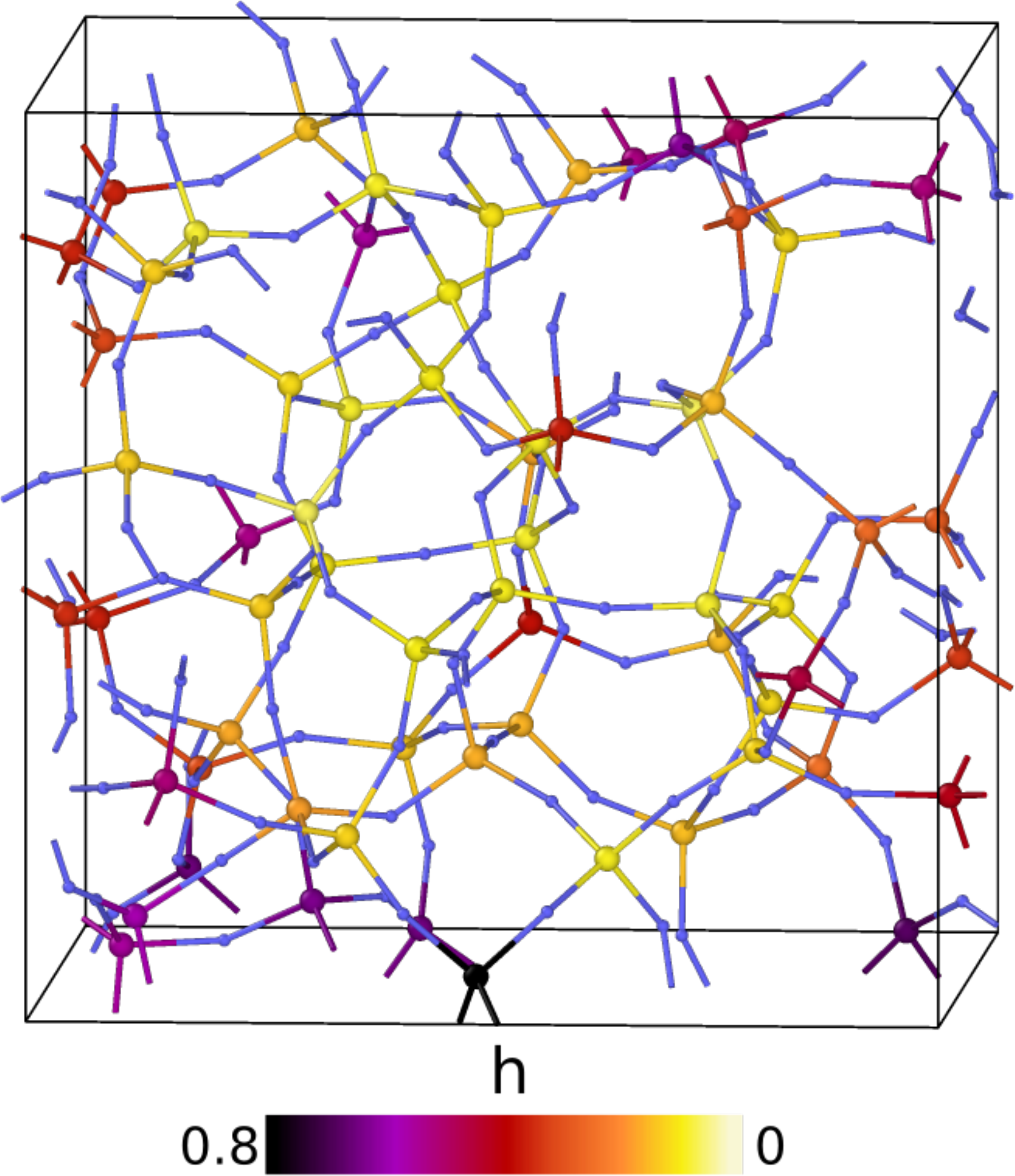}
    b)
    \includegraphics[width=0.4\textwidth]{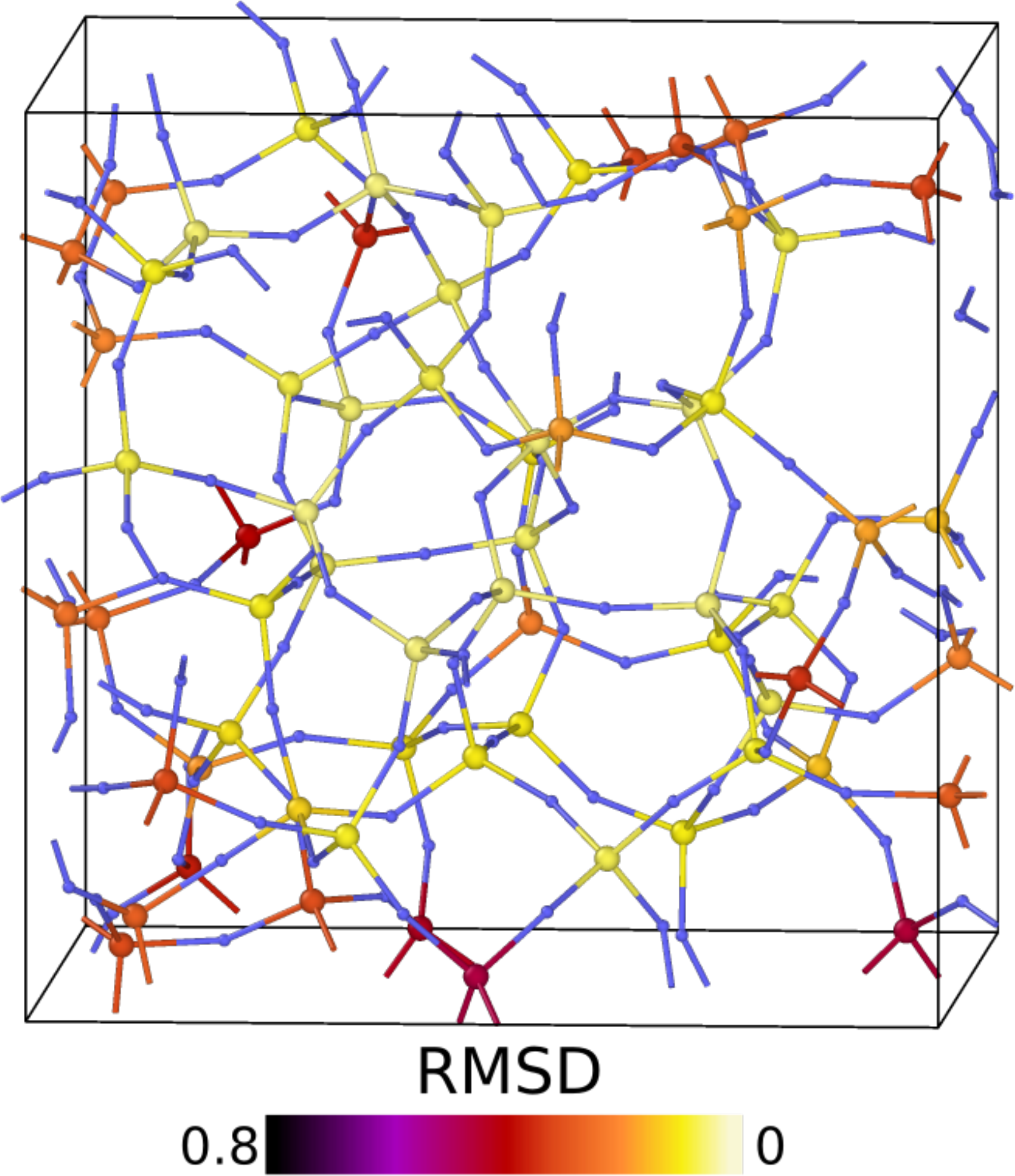}
    c)
    \includegraphics[width=0.4\linewidth]{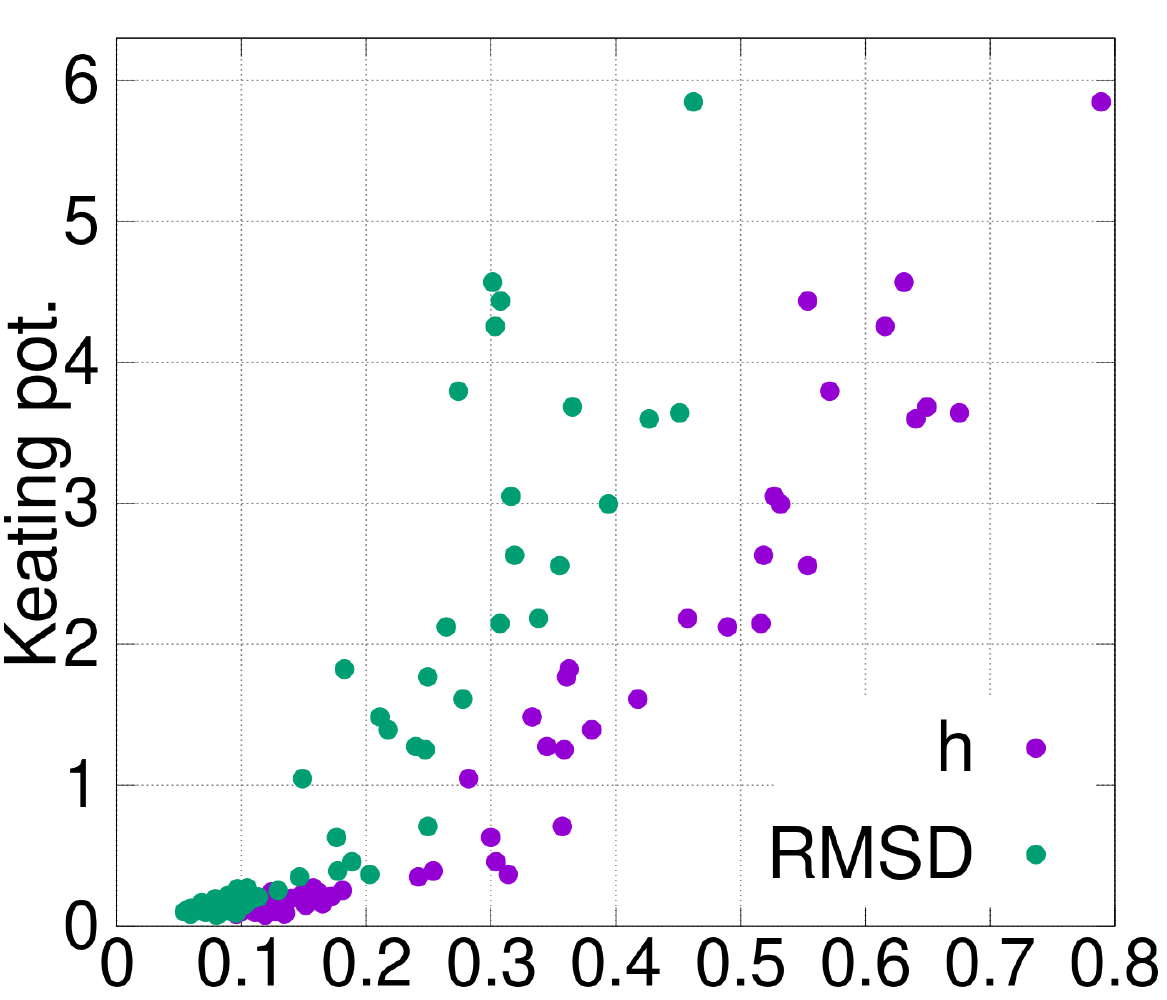}
    \caption{Color map of distortions in a 192 atoms silica model, as obtained through a) $h(A,B)$, b) $RMSD(A,B)$, and c) correlation with respect to Keating potential \cite{Pot_KT}.}
    \label{fig:asi_d}
\end{figure*}

In the context of amorphous or disordered structures, it can also enable the characterization and analysis of local disorder at different scales, i.e. as a function of the number of neighbors included in the fragment and accounted during matching. Fig.~\ref{fig:asi_d}, shows the Hausdorff and $RMSD$ distance color map for SiO$_4$ tetrahedra in silica. 
In this example, IRA was used to find the matching between an ideal SiO$_4$ tetrahedron and the whole silica crystal, centered on each of the Si atoms.
The O atoms are shown in blue, and Si atoms are colored by the value of chosen distance function. The color map is compared to the values obtained through the Keating potential\cite{Pot_KT}, which is a strain-based potential, where a low value corresponds to Si atoms with local environments closely resembling a tetrahedron (low strain), and higher values otherwise (higher strain).

Finally, because of the ability of our approach to match non-connected fragments, it can be also exploited to compute time correlation functions based on fragments taken at two different times.

\section{Conclusion}
In this work, we have presented an alternative, parameter-less shape matching approach that allows to find isometric transformations (rigid rotation, reflection, translation, and permutation/atomic assignment) between congruent and near-congruent structures that do not necessarily have the same number of atoms, and that can be part of a periodic lattice. 
The {\it best match} transformation coincides with a minimum of the set-set distance, which has value zero in case of exact congruence between the structures.
As such, the set-set distance can be interpreted as a measure of similarity, thus enabling the use of our approach for comparing and recognizing atomic structures. 
The CShDA algorithm, the LAP solver we developed, is able to compute atomic assignments for structures with non-equal number of atoms. This is exploited in the IRA algorithm, and enables the resolution of the shape matching problem for structural fragments.
Among the performed tests, the reliability of the algorithm is 100\% in the case of exact congruence of structures (Sec.~\ref{sec:tests_cdb}), while the performances might drop slightly for larger deformations (99.6\% in the cyanine case Sec.~\ref{sec:cyanine}). When available, prior knowledge of the structures can be exploited to reduce the number of mismatches.
In the context of finding correlations and identifying collective behaviours, our approach could aid in revealing the most important collective axes, either in space or time.

\section*{Data and source code availability}
IRA is released under double licensing, GPL v3 and Apache v2. The source code and data used for testing and benchmarking is available at \url{https://github.com/mammasmias/IterativeRotationsAssignments}. For cyanine trajectory please contact authors in Ref.~\cite{sara_cyanine}.

\vspace{0.2cm}
{\bf ACKNOWLEDGEMENT}
The authors are active members of the Multiscale And Multi-Model Approach for MaterialS In Applied Science consortium (MAMMASMIAS consortium), and acknowledge the efforts of the consortium in fostering scientific collaboration. 
This work was partially funded from the European Union’s Horizon 2020 research and innovation program under grant agreement No. 871813 MUNDFAB, and by the European Union’s Horizon 2020 research and innovation program under grant agreement No 899285 MAGNELIQ. 
All images of atomic structures in this article were generated with ovito\cite{ovito} software.

{\bf ASSOCIATED CONTENT:}
Supporting Information is Available free of charges. The supporting Information contains detailed figures of the benchmark test results as well as a detailed analysis on the rotations that are required to find the approximate rotation transformation.

\bibliographystyle{apsrev4-1}
\bibliography{biblio_short}

\appendix
\clearpage

\setcounter{table}{0}
\renewcommand{\thetable}{S\arabic{table}}%
\setcounter{figure}{0}
\renewcommand{\thefigure}{S\arabic{figure}}%

\section*{ SUPPORTING INFORMATION}

\subsection{Results of the benchmark test}
\begin{figure*}
    \centering
    \includegraphics[width=0.8\linewidth]{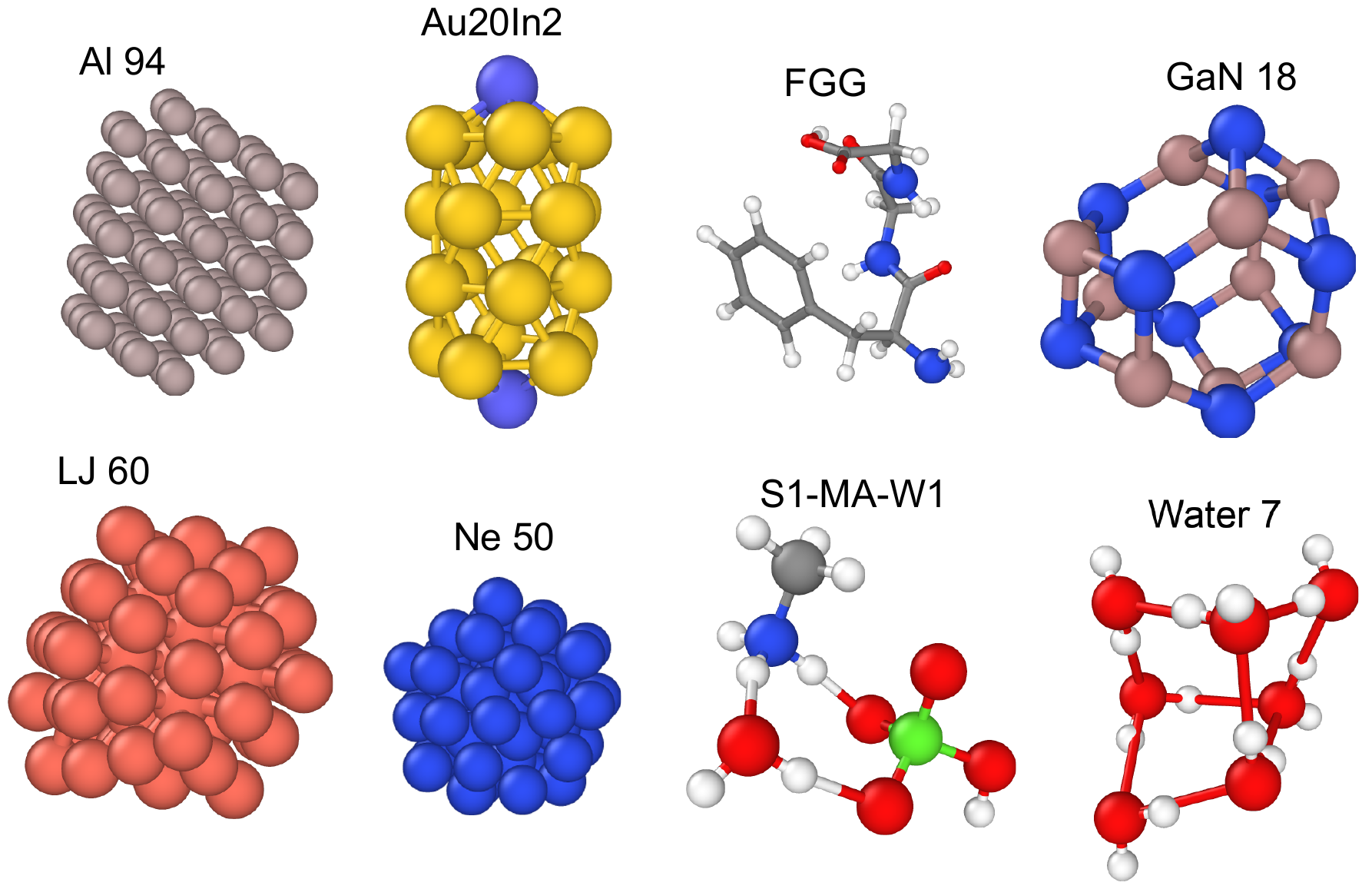}
    \caption{Representative structures from each dataset used in the benchmark test of Sec.~\ref{sec:tests_cdb}. Note the diversity of general shape in the structures.}
    \label{fig:bench_structures}
\end{figure*}

The Fig.~\ref{fig:bench_structures} shows representative structures from each dataset included in the benchmark test of Sec.~\ref{sec:tests_cdb}. The collection of structures included in the benchmark test forms a diverse set of general shapes. More details about these structures can be found in their respective original works\cite{al_dataset, GaN_dataset, au26_dataset, cdb, temelso2017}.

\begin{figure*}
    \centering
    \includegraphics[width=0.7\linewidth]{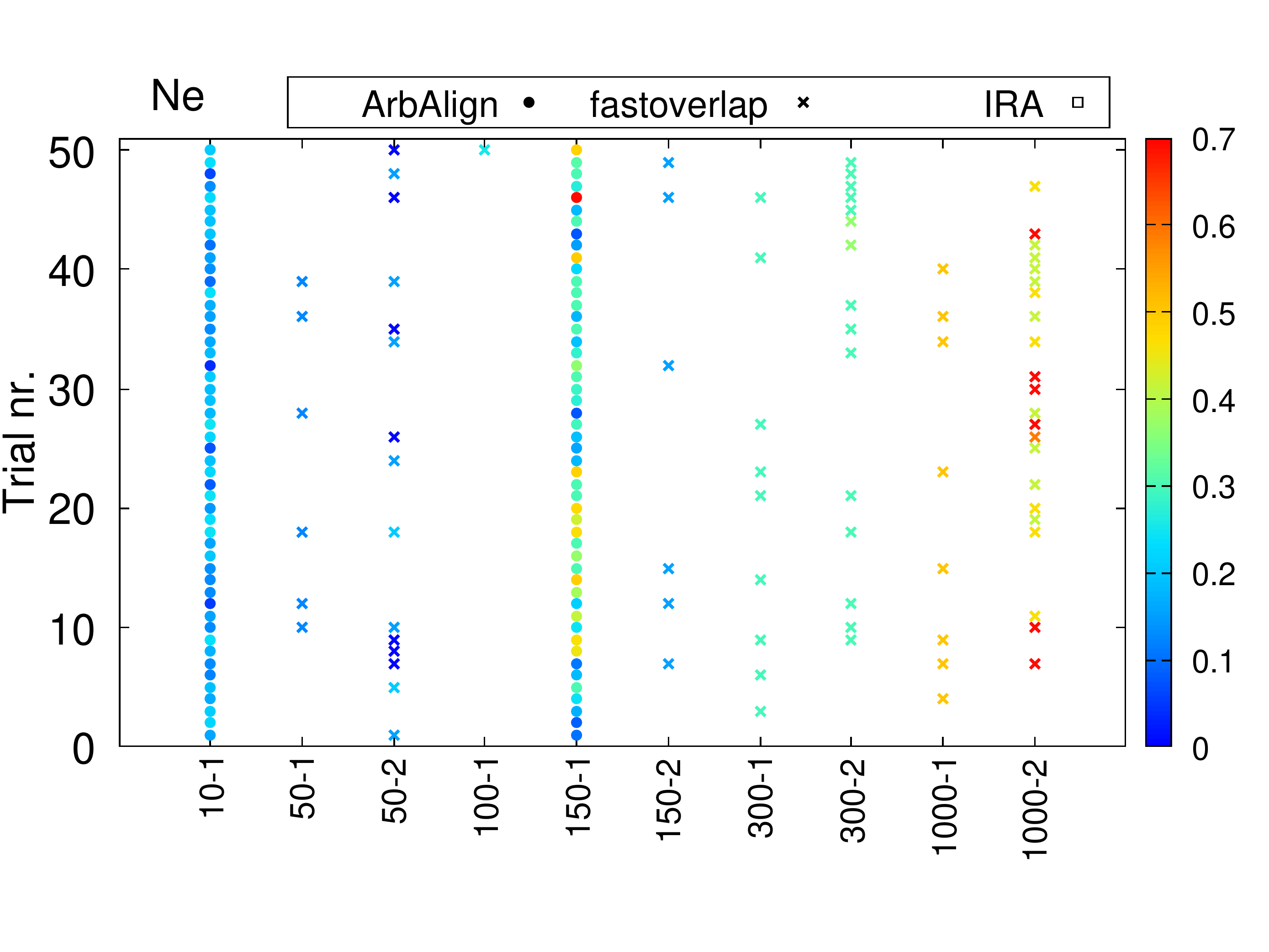}
    \caption{Values of final $RMSD$ for structures from the Ne dataset. Only failures are reported. Structure name on horizontal axis, trial number on vertical, final $RMSD$ value in color. Failures in this dataset: 100 failures in 2 structures by ArbAlign; 82 failures in 8 structures by fastoverlap; 0 failures by IRA.}
    \label{fig:pl_Ne}
\end{figure*}

\begin{figure*}
    \centering
    \includegraphics[width=0.5\linewidth]{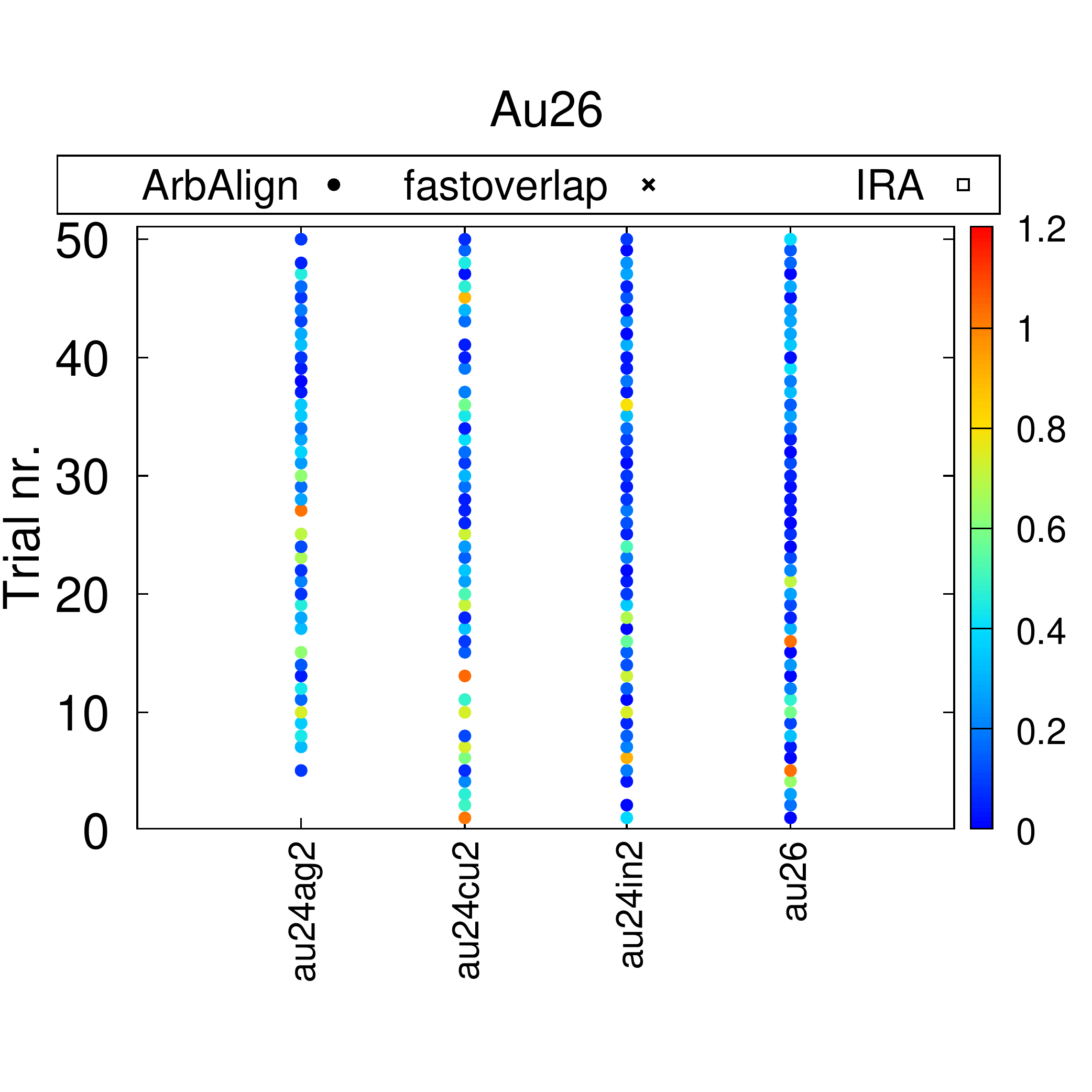}
    \caption{Values of final $RMSD$ for structures from the Au26 dataset. Only failures are reported. Structure name on horizontal axis, trial number on vertical, final $RMSD$ value in color. Failures in this dataset: 186 failures in 4 structures by ArbAlign; 0 failures by fastoverlap; 0 failures by IRA.}
    \label{fig:pl_Au}
\end{figure*}

\begin{figure*}
    \centering
    \includegraphics[width=0.7\linewidth]{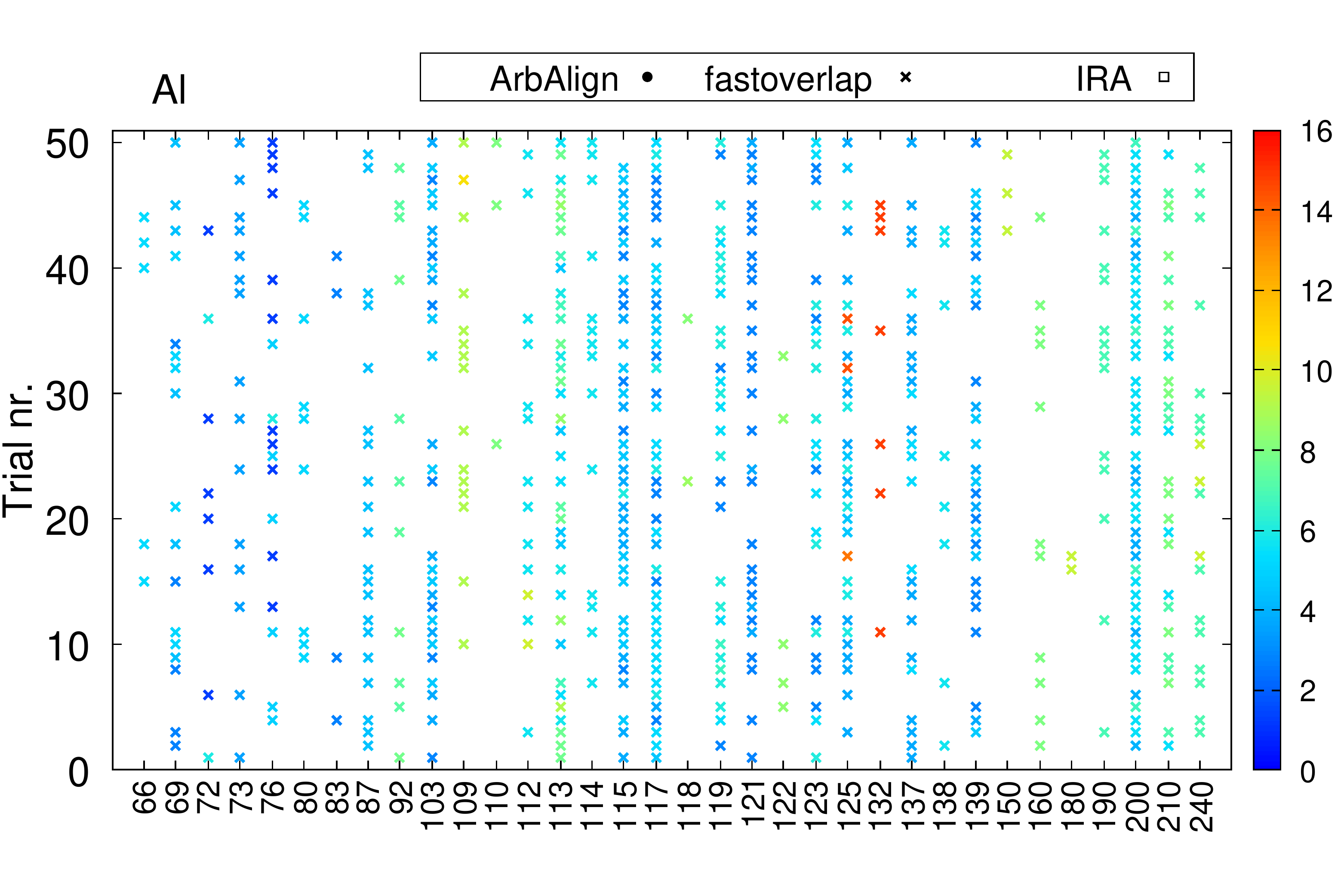}
    \caption{Values of final $RMSD$ for structures from the Al dataset. Only failures are reported. Structure name on horizontal axis, trial number on vertical, final $RMSD$ value in color. Failures in this dataset: 0 failures by ArbAlign; 613 failures in 34 structures by fastoverlap; 0 failures by IRA.}
    \label{fig:pl_Al}
\end{figure*}

\begin{figure*}
    \centering
    \includegraphics[width=0.7\linewidth]{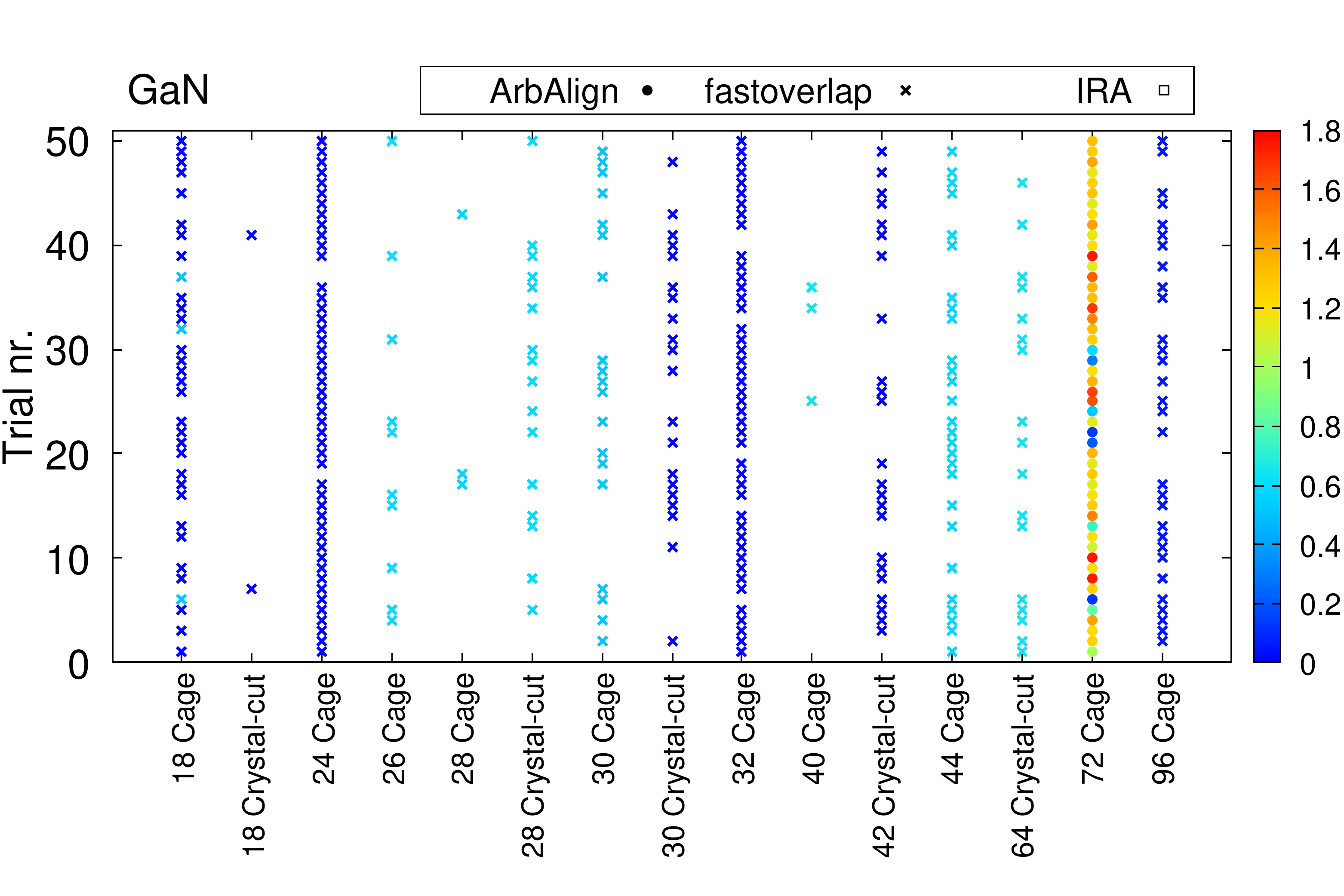}
    \caption{Values of final $RMSD$ for structures from the GaN dataset. Only failures are reported. Structure name on horizontal axis, trial number on vertical, final $RMSD$ value in color. Failures in this dataset: 50 failures in 1 structure by ArbAlign; 294 failures in 14 structures by fastoverlap; 0 failures by IRA.}
    \label{fig:pl_GaN}
\end{figure*}

\begin{figure*}
    \centering
    \includegraphics[width=0.7\linewidth]{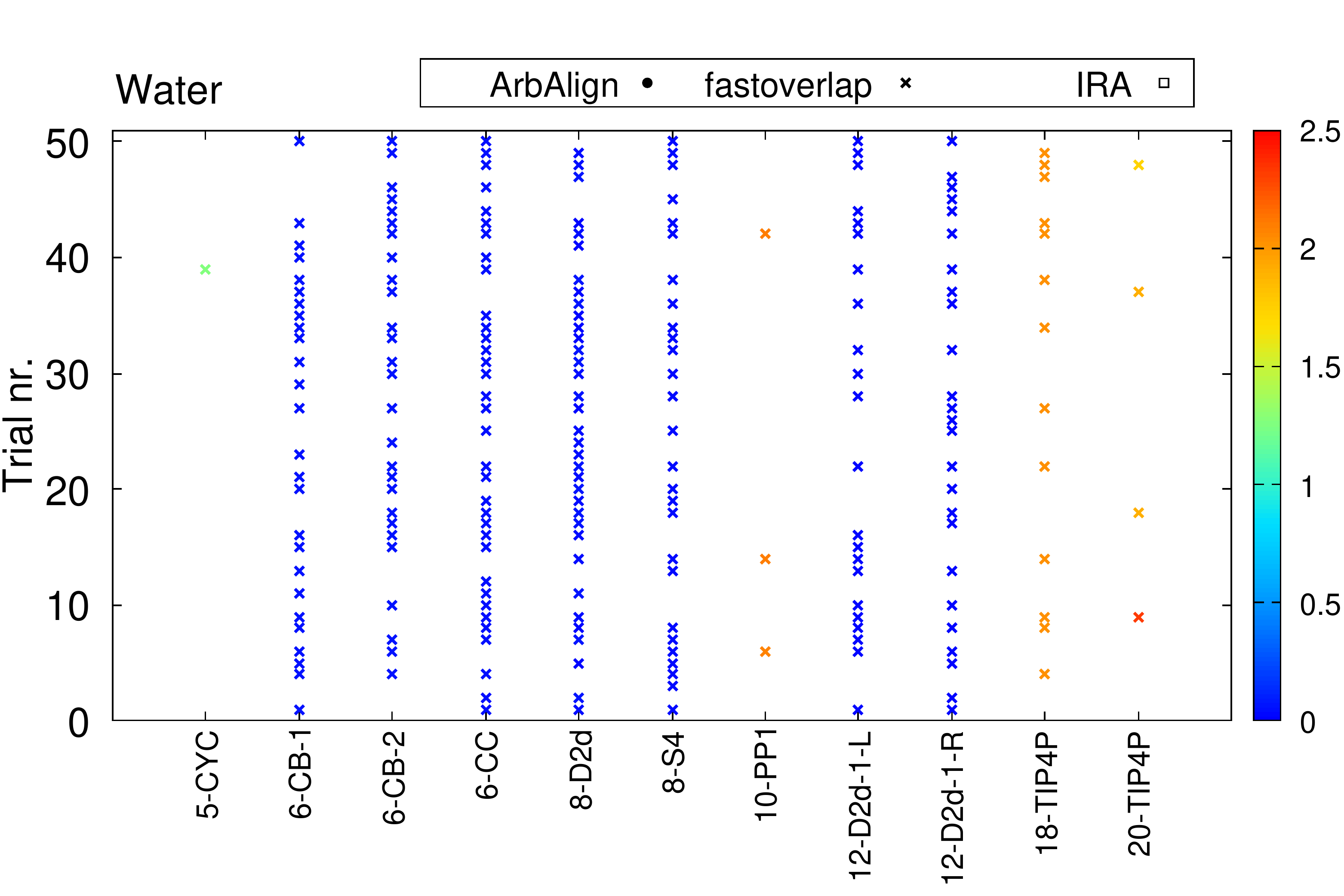}
    \caption{Values of final $RMSD$ for structures from the water dataset. Only failures are reported. Structure name on horizontal axis, trial number on vertical, final $RMSD$ value in color. Failures in this dataset: 0 failures by ArbAlign; 217 failures in 11 structures by fastoverlap; 0 failures by IRA.}
    \label{fig:pl_water}
\end{figure*}

\begin{figure*}
    \centering
    \includegraphics[width=\linewidth]{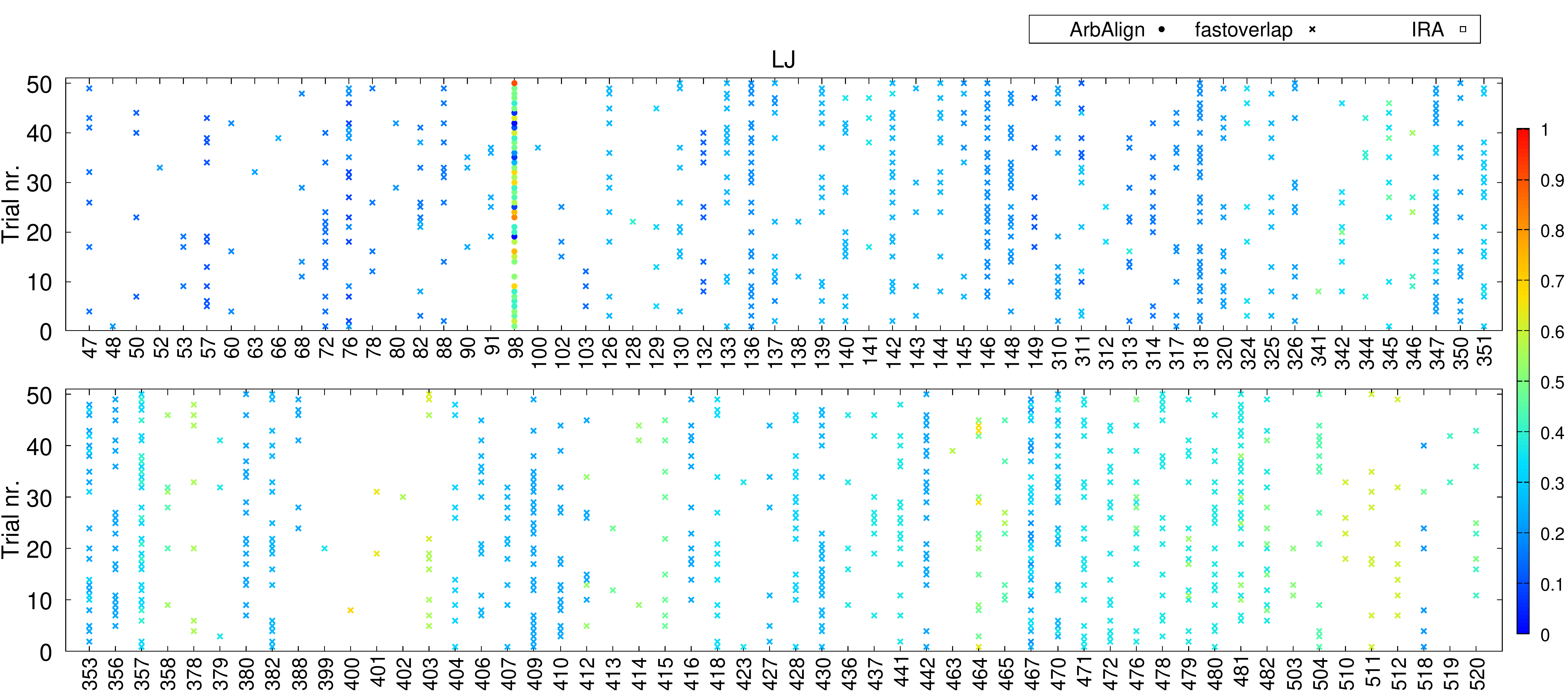}
    \caption{Values of final $RMSD$ for structures from the LJ dataset. Only failures are reported. Structure name on horizontal axis, trial number on vertical, final $RMSD$ value in color. Failures in this dataset: 45 failures in 1 structure by ArbAlign; 1177 failures in 113 structures by fastoverlap; 0 failures by IRA.}
    \label{fig:pl_LJ}
\end{figure*}

A final transformation having $RMSD(A,B)>0.001$ is considered a mismatch. Failures are reported for each software in Figs.~\ref{fig:pl_Ne}-\ref{fig:pl_LJ}. The horizontal axis on these plots gives the name of the particular structure where a failure has occurred, the vertical axis is the number of current trial, the color of a point gives the final value $RMSD$, and the shape of a point is related to the particular software which returned the failure.


\subsection{Number of rotations tested}
\begin{figure*}[h!]
    \centering
    \includegraphics[width=0.8\textwidth]{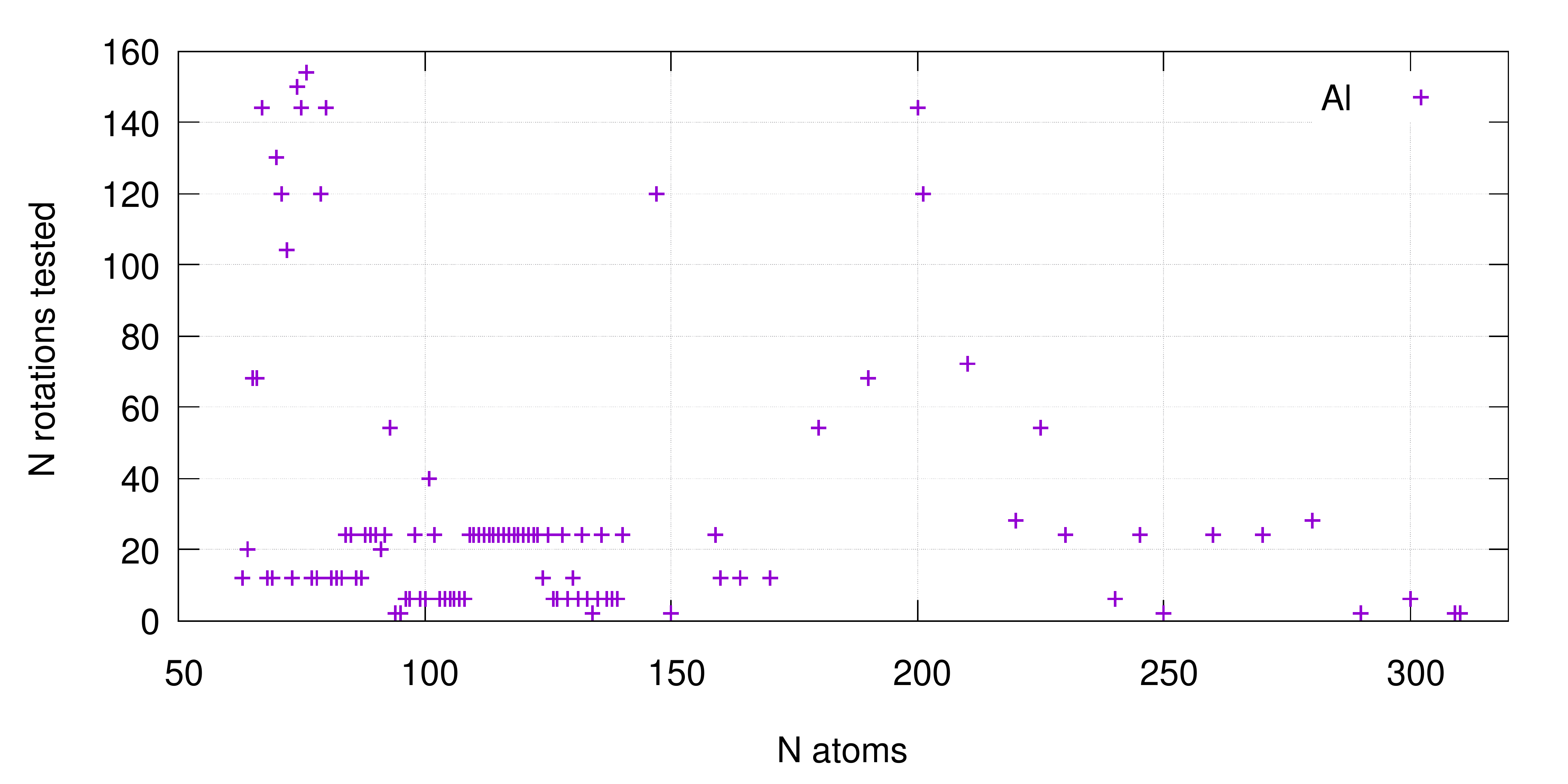}
    \caption{Number of rotations tested versus the number of atoms, for structures in the Al dataset \cite{al_dataset}. }
    \label{fig:al_nbas}
\end{figure*}

Fig.~\ref{fig:al_nbas} shows the number of rotations tested for all structures in the Al dataset, versus the total number of atoms in the structure.
As it can be seen, the number of rotations tested is on the range [2, 154] and there is no apparent rule. The number of tested reference frames is related to the structure surrounding the origin point as mentioned in Sec.~\ref{sec:guess}, which in the case of non-equal number of atoms is a central atom, and in the case of equal number of atoms is the geometrical center (or any known common point). The higher number of tested rotations occurs when the geometrical center of the structure coincides with an atomic position. In that case, the distance to nearest atoms is the highest. A large number of atoms is therefore included in the radial cutoff region, such increasing the number of possible reference frames to be tested. When the geometrical center falls in between atoms, the distance to nearest neighbors is shorter (lower number of atoms), and thus less reference frames have to be tested.

\end{document}